\documentclass[prb,twocolumn,floatfix]{revtex4}

\usepackage{amsfonts}
\usepackage{amsmath}
\usepackage{verbatim}
\usepackage[dvips]{graphicx}
\usepackage{subfigure}

\def\Tr{\textrm}
\def\ra{\rangle}            
\def\la{\langle}            
\def\dd{\textrm{d}}
\def\Bf{\boldsymbol}

\begin{document}

\title{Effective action for vortex dynamics in clean $d$-wave superconductors}
\author{Predrag Nikoli\'c and Subir Sachdev}
\affiliation{Department of Physics, Harvard University, Cambridge MA
02138, USA}
\date{November 11, 2005}
\pacs{}

\begin{abstract}
We describe the influence of the gapless, nodal, fermionic
quasiparticles of a two-dimensional $d$-wave superconductor on the
motion of vortices. A continuum, functional formalism is used to
obtain the effective vortex action, after the fermions have been
integrated out. At zero temperture ($T$), the leading terms in the
vortex action retain their original form, with only a finite
renormalization of the vortex effective mass from the fermions. A
universal ``sub-Ohmic'' damping of the vortex motion is also found.
At $T>0$, we find a Bardeen-Stephen viscous drag term, with a
universal co-efficient which vanishes as $\sim T^2$. We present a
simple scaling interpretation of our results, in which
quantum-critical Dirac fermions respond to a moving point
singularity. Our results appear to differ from those of the
semiclassical theory, which obtains more singular corrections to a
vortex mass appearing in transport equations.
\end{abstract}

\maketitle

\section{Introduction}
\label{sec:intro}

The cuprate superconductors behave like conventional BCS
superconductors in many respects, albeit with a $d$-wave symmetry
of the Cooper pair wavefunction. The BCS wavefunction provides a
reasonable description of the ground state. The elementary
excitations above the ground state are \\
({\rm i\/}) fermionic, $S=1/2$, Bogoliubov quasiparticles, \\({\em
ii\/}) plasmons (or ``phase fluctuations''), and\\ ({\em iii\/})
vortices with magnetic flux $hc/(2e)$.

For the traditional ``low temperature'' superconductors, the
vortices are usually treated as a classical, configuration of the
superconducting order parameter, which forms a background for the
quantum fluctuations of the fermionic quasiparticles and the Cooper
pairs: the quantum zero-point motion of the vortices themselves is
negligibly small.

Quantum fluctuations of the vortices are expected to play a more
fundamental role in the cuprate superconductors, requiring that
the vortices be treated as bona fide quasiparticles and elementary
excitations in their own right. There are a variety of
experimental indications
\cite{ando,fang,ali,mcelroy,hanaguri,steiner,tranquada} that the
superconducting state is proximate to an insulating ground state.
Vortices and anti-vortices are expected to proliferate in the
vicinity of such a transition, and so the energy gap towards
creation of vortex/anti-vortex pairs in the superconductor becomes
vanishingly small as the superconductor-insulator transition is
approached. Moreover, the superconductor coherence length, $\xi$,
is at most a few lattice spacings, which suggests that the
effective mass of the vortices is small.

\begin{figure}
\includegraphics[width=2.5in]{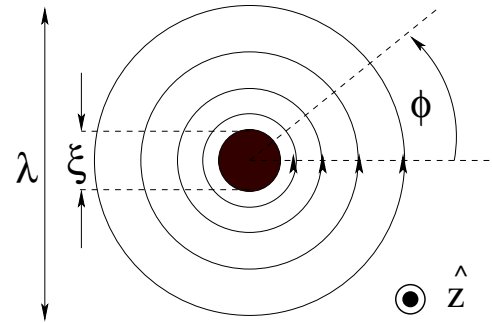}
\caption{\label{vortex}Scales of a vortex. The core size is of the
order of superconducting coherence length $\xi$, while the extent
of supercurrent circulations is roughly equal to the much larger
London penetration depth $\lambda$. This paper will work in the
limit in which $\lambda \rightarrow \infty$.}
\end{figure}

It is the purpose of this paper to discuss some features of the
effective action controlling interplay between the three classes of
elementary excitations (noted above) of a two-dimensional $d$-wave
superconductor. If we neglect the fermionic $S=1/2$ quasiparticles,
the required action can be readily deduced from the theory of
boson-vortex duality\cite{dh,nelson,fisherlee}, as we will review
below. We will primarily be concerned here on the influence of the
gapless, nodal, fermionic quasiparticles on the vortices. The
interaction between the fermions and the vortices is non-local, with
no natural expansion parameter, and so its description poses a
strong-coupling problem. Here, we will consider a single, isolated,
slowly moving vortex, and exactly evaluate the functional integral
over the fermionic degrees of freedom by a combination of analytic
and numerical techniques; this yields a renormalization of the
effective action for the vortices and the plasmons. Our primary
result will be that despite the absence of an energy gap in the
fermionic spectrum, the influence of the fermionic degrees of
freedom is relatively innocuous, and at zero temperature ($T$) the
leading terms in the vortex action retain the form deduced while
ignoring the fermionic quasiparticles.

The interplay between fermions and moving vortices has also been
examined previously in semiclassical framework by a number of
workers.\cite{volovik,KopVin,KopVin2,kopnin} These theories are
expansion in the ratio of the superconducting pairing energy to the
Fermi energy, a ratio which is not small in the cuprate
superconductors. We will not make any assumptions on the value of
this ratio in the present paper. The results of semiclassical theory
are expressed in a different framework of transport equations, and
this makes direct comparison with our results difficult. An
important characteristic of the semiclassical results is that the
effective mass of a vortex (as defined in their transport equations)
in a $d$-wave superconductor diverges linearly with the spacing
between vortices. We do not find any corresponding divergence in the
vortex mass in our functional formalism. The reasons for this
potential discrepancy between our and the semiclassical results are
not completely clear to us. We speculate it may be due to the
breakdown of the semiclassical approximation in describing the low
energy nodal fermionic quasiparticles. Such fermionic excitations
are not semiclassical, but, with their scale-invariant Dirac
spectrum, better characterized as a quantum-critical
system.\cite{sl}

We begin our presentation by reviewing the action for vortices and
plasmons as derived in theories of the boson-vortex
duality,\cite{dh,nelson,fisherlee} while ignoring fermionic
excitations of the superconductor. We use a first-quantized
approach for the vortex degrees of freedom, and treat each vortex
as a quantum particle at spacetime position $\Bf{r}_{\Tr{v}} (
\tau)$, where $\tau$ is imaginary time; the quantum mechanics of
the vortices is described by a functional integral over
$\Bf{r}_{\Tr{v}} ( \tau)$. The density or ``phase'' fluctuations
of the superconductor are represented by a U(1) gauge field
$\mathcal{A}_\mu (\Bf{r}, \tau)$, where the spacetime index $\mu$
extends over $x$, $y$, $\tau$: the `magnetic' flux of this gauge
field is proportional to the particle density of the
superconductor, while the `electric' field determines the local
superflow velocity. A vortex (anti-vortex) is minimally coupled to
the U(1) gauge fields with charge $+1$ ($-1$). So a single vortex
and the phase fluctuations are described by the action
$\mathcal{S} = \mathcal{S}_v + \mathcal{S}_{\mathcal{A}}$ where
\begin{eqnarray}
\mathcal{S}_v &=& \int d\tau \Biggl[ \frac{m_{\Tr{v}}^0}{2} \left(
\frac{d \Bf{r}_{\Tr{v}} ( \tau)}{d \tau } \right)^2 + i \frac{d
\Bf{r}_{\Tr{v}} ( \tau)}{d \tau } \cdot \vec{\mathcal{A}}^0
(\Bf{r}_{\Tr{v}} ( \tau)) \nonumber \\
&~&~~~~~~~~~~+ i \int d^2 \Bf{r} d \tau \mathcal{A}_{\mu} (\Bf{r},
\tau) J_{v\mu} (\Bf{r}, \tau) \Biggr] \label{sv}
\end{eqnarray}
is the vortex action and its coupling to the U(1) gauge field. Here
$m_{\Tr{v}}^0$ is the bare vortex mass (which can be viewed as
arising from high energy degrees of freedom which have been
integrated out), $\vec{\mathcal{A}}^0$ represents the `background'
flux which is responsible for the Magnus force on the vortices. In a
Galilean-invariant superfluid, $\nabla \times \vec{\mathcal{A}}^0 =
\hat{\Bf{z}} 2 \pi (\mbox{density of Cooper pairs})$. In the
presence of a lattice, when the superconductor is proximate to a
Mott insulator, it has been
argued\cite{CompOrd1,CompOrd2,EstVMass,CompOrdRev} that the Magnus
force is reduced, and the density of pairs in the Mott insulator has
to be subtracted from the density of Cooper pairs. In the last terms
in $\mathcal{S}$, $J_{v\mu}$ is the vortex 3-current, and its
temporal and spatial components are given by
\begin{eqnarray}
J_{v\tau} (\Bf{r}, \tau) &=& \delta ( \Bf{r} - \Bf{r}_{\Tr{v}}
(\tau))
\nonumber \\
\Bf{J}_{v} (\Bf{r}, \tau) &=& \delta ( \Bf{r} - \Bf{r}_{\Tr{v}}
(\tau)) \frac{d \Bf{r}_{\Tr{v}} ( \tau)}{d \tau }
\end{eqnarray}
The phase fluctuations of the superconductor are controlled by the
action $\mathcal{S}_\mathcal{A}$ which, in the Coulomb gauge $\nabla
\cdot \vec{\mathcal{A}} = 0$, is
\begin{eqnarray}
\mathcal{S}_\mathcal{A} &=& \int \frac{d^2 k d \omega}{8 \pi^3}
\Biggl( \frac{1}{8 \pi^2 \rho_s} \left[ k^2 |\mathcal{A}_\tau (k,
\omega) |^2 +
\omega^2 | \mathcal{A}_{i} (k, \omega) |^2 \right] \nonumber \\
&~&~~~+ \frac{e^{\ast 2} k}{4\pi} \left[ \delta_{ij} - \frac{k_i
k_j}{k^2} \right] \mathcal{A}_i (-k,-\omega) \mathcal{A}_{j} (k,
\omega) \Biggr). \label{sa}
\end{eqnarray}
We have Fourier transformed to momentum, $k$, and imaginary
frequencies, $\omega$, $e^\ast = 2e$ is the charge of a Cooper
pair, and $\rho_s$ is the superfluid stiffness in units of energy
(the London penetration depth, $\lambda$,  is related to $\rho_s$
by $\rho_s = \hbar^2 c^2 d/(16 \pi e^2 \lambda^2)$, where $d$ is
the interlayer spacing). The first term in
$\mathcal{S}_\mathcal{A}$ is responsible for the log interaction
between the vortices; the coefficient of this term is fixed by
requiring that the log interaction have the required prefactor.
The second term represents the electrical Coulomb interaction
between the charged Cooper pairs; its coefficient is fixed by the
knowledge that $(\nabla \times \vec{\mathcal{A}})/(2 \pi)$ is the
fluctuation in the number density of pairs of charge $e^\ast$. The
`photon' excitations of the full action $\mathcal{S}_\mathcal{A}$
are the plasmons of the two-dimensional superconductor with
dispersion $\omega \sim \sqrt{k}$, and these plasmons can be
`radiated' by a moving vortex.

It is worth pointing out here a noteworthy aspect of the action for
vortices and plasmons presented here in Eqs.~(\ref{sv}) and
(\ref{sa}). Although obtained by an approximate and uncontrolled
boson-vortex duality mapping, most of the coupling constants in the
final action can be fixed rather precisely. The primary unknown is
the vortex mass, $m_{\Tr{v}}$, and we will describe microscopic
contributions to its value this paper.

Clearly, the physics described by $\mathcal{S}_v +
\mathcal{S}_\mathcal{A}$ is similar to the quantum electrodynamics
of a charged particle. The coupling to the phase fluctuations will
renormalize the vortex mass, $m_{\Tr{v}}$, and lead to `radiation
damping' in the motion of a vortex. We review this physics in
Section~\ref{sec:dynamic}. For a neutral superfluid, it has been
argued\cite{popov,duan} that the phase fluctuations lead to a
logarithmically divergent ``dynamic mass'' of the vortex; we show
how the analog of this result obtains in our formalism in
Section~\ref{sec:dynamic}. Section~\ref{sec:dynamic} also considers
the case of a charged superfluid, with plasmon excitations as in
Eq.~(\ref{sa}): for this case, we find that the phase fluctuations
are less singular and lead only to a {\em finite\/} renormalization
of the vortex ``dynamic mass''.

Our purpose here is to extend $\mathcal{S}_v +
\mathcal{S}_\mathcal{A}$ to include the remaining elementary
excitations of the $d$-wave superconductor, the nodal fermionic
quasiparticles. (The complete action controlling the interactions
between all the elementary excitations of a $d$-wave superconductor
is \cite{sf,bf,grsf,bs} $\widetilde{\mathcal{S}}_v +
\mathcal{S}_\mathcal{A} + \mathcal{S}_\Psi + \mathcal{S}_{cs}$
specified in Eqs.~(\ref{svf}), (\ref{sa}), (\ref{spsif}), and
(\ref{scs})). We will then integrate out these quasiparticles, and
associated gauge fields (including $\mathcal{A}_\mu$), in the
presence of a single, slowly moving vortex, and obtain resulting
renormalization of $\mathcal{S}_v$. For small $\Bf{r}_{\Tr{v}}
(\omega)$, the effective vortex action can be written in the form
(apart from terms responsible for the Magnus force associated with
$\vec{\mathcal{A}}^0$ in Eq.~(\ref{sv}))
\begin{equation}
\widetilde{\mathcal{S}}_{v} = \int \frac{d \omega}{2 \pi}
F_{\parallel} (\omega) |\Bf{r}_{\Tr{v}} (\omega) |^2 + \ldots
\label{tsv}
\end{equation}
For small $\omega$, we find for a charged $d$-wave superconductor at
$T=0$ that the contribution of the nodal quasiparticles to
$F_{\parallel} (\omega)$ can be written as
\begin{equation}
v_F^2 F_{\parallel} (\omega ) = \omega^2 \Lambda F_1 ( v_\Delta/v_F)
- |\omega|^3 F_2 (v_\Delta / v_F), \label{nonohmic}
\end{equation}
where $v_F$ and $v_\Delta$ are the two velocities of the nodal
quasiparticles, $\Lambda$ is a high energy cutoff, and $F_{1,2}$ are
functions of the ratio of the velocities. Comparing with
Eq.~(\ref{sv}), we see that the first term can be interpreted as
finite renormalized mass $m_{\Tr{v}} = \Lambda F_1 (v_\Delta/v_F)/(2
v_F^2 )$ of the vortex. The function $F_1$ is not universal, and
should clearly depend upon the nature of the cutoff; however, no
other coupling constant appears in the value of $F_1$. The second
$|\omega|^3$ term in Eq.~(\ref{nonohmic}) represents a ``sub-Ohmic''
damping of the vortex motion. We will show that this damping has a
{\em universal\/} strength {\em i.e.\/} $F_2$ is a universal
function of $v_\Delta/v_F$, and so the damping is entirely
determined by the two quasiparticle velocities. (For completeness,
we note that in a neutral $d$-wave superfluid, as we will
demonstrate in Section~\ref{sec:dynamic}, there is an additional
$\omega^2 \ln (\Lambda/|\omega|)$ term in $F_{\parallel} (\omega)$:
this term is the manifestation, in the present formalism, of the
logarithmically divergent ``dynamic mass'' of the vortex
\cite{popov,duan}, noted above.)

It is interesting to note here that the expression for
$F_{\parallel} (\omega)$ in Eq.~(\ref{nonohmic}) could have been
deduced very simply and directly by dimensional arguments, after
assuming that the nodal fermions are properly thought of as a
`quantum critical' system with dynamic exponent $z=1$, which then
responds to the moving vortex. In this scaling argument, the vortex
position $\Bf{r}_{\Tr{v}} (\tau)$ scales as a length, and so
$\mbox{dim} [\Bf{r}_{\Tr{v}} (\tau)]=-1$. After a Fourier
transformation, $\mbox{dim} [\Bf{r}_{\Tr{v}} (\omega)]=-2$. Counting
dimensions in the action, we conclude then that $\mbox{dim} [
F_\parallel (\omega)] = 3$. Also, we know from translational
invariance that $F_\parallel (0) = 0$. So apart from a universal
$|\omega|^3$ term, $F_\parallel (\omega)$ can have a
cutoff-dependent, analytic $\omega^2$ term. The dependence of
$F_\parallel (\omega)$ on the velocities follows from ensuring
consistency with engineering dimensions. The success of this
argument demonstrates the importance of thinking about the fermions
in terms of the scale-invariant critical field theory that obeys
hyperscaling properties. As we noted earlier, we believe that it is
this criticality which is not properly captured by the semiclassical
analysis.\cite{volovik,KopVin,KopVin2,kopnin}

This scaling argument, and our explicit computations, show that no
$|\omega|$ ``Ohmic'' damping term is generated in $F_\parallel
(\omega)$ at $T=0$. If present, such a term would contribute the
familiar Bardeen-Stephen damping\cite{bardeen} to the vortex
equations of motion. If the vortex mass had the degree of infrared
divergence claimed in the semiclassical
theory\cite{volovik,KopVin,KopVin2,kopnin} {\em i.e.\/} a mass which
diverges linearly with length scale, then this would be equivalent,
in our language, to a scaling dimension $\mbox{dim}[F_\parallel
(\omega)] = 1$, and would imply that a $|\omega|$ term should have
appeared in our vortex action. However, because we find
$\mbox{dim}[F_\parallel (\omega) ] = 3$, all our integrals have
a degree of convergence in the infrared to allow safe expansion in
powers of $\omega$ upto order $\omega^2$, and to prevent the
appearance of a $|\omega|$ term.

In the real experimental system, our results show that any
Bardeen-Stephen damping must come from other sources: a non-zero
$T$, or from impurities which induce a non-zero density of fermionic
states at zero energy. In particular, we show in
Section~\ref{ssecFiniteT} that at $T>0$, the function $F_{\parallel}
(\omega)$ contains a Bardeen-Stephen term
\begin{equation}
F_\parallel (\omega) = \frac{\eta}{2} |\omega| + \ldots
\end{equation}
where the viscous drag co-efficient is universal in the clean limit:
\begin{equation}
\eta = \frac{(k_B T)^2}{v_F^2} F_\eta (v_\Delta/v_F),
\end{equation}
with $F_\eta$ a universal function of the velocity ratio. We are
able to compute one contribution to $\eta$ exactly: that due to the
statistical phase factor acquired by the fermions upon encircling
the vortex---the result appears in Eq.~(\ref{FPT}). There is a
second contribution due to the Doppler shift which requires a
numerical analysis.

We will begin by setting up the general formalism
\cite{Simanek,AoZhu2,AoZhu,Han} under which the effective action of
a single vortex can be derived in Section~\ref{secVortexAction}. The
nature of the coupling between the fermionic quasiparticles and the
vortices and the plasmons will then be reviewed in
Section~\ref{secHamiltonian}; here we will follow the framework of
Te\v sanovi\'c and others.\cite{ft,Tesh2,Tesh1,halperin,ashwin}

With the basic formalism in place, it then remains to explicitly
evaluate the functional determinant of the nodal fermions. There is
no natural small parameter in this determinant, and so an exact
evaluation is required. Nevertheless, it is useful to examine the
structure of the perturbation theory order-by-order in the
non-linearities; this will be done in
Section~\ref{sec:perturbation}. We will find that the perturbation
theory has the scaling structure discussed near
Eq.~(\ref{nonohmic}).

The non-perturbative evaluation of the fermion determinant is
presented in Sections~\ref{secAnalytic} and~\ref{secNumerics}. In
Section~\ref{secAnalytic} we neglect the `Doppler shift' of the
fermion energy in the presence of a superflow, and retain only the
`Berry phase' acquired by the fermions upon encircling a vortex; we
show that an analytic evaluation of the effective mass
renormalization is possible in this limit. Section~\ref{secNumerics}
accounts for the Doppler shift in a numerical computation. These
non-perturbative results are found to be entirely consistent with
the expectation of the perturbation theory discussed in
Section~\ref{sec:perturbation} in that they also has the scaling
structure of Eq.~(\ref{nonohmic}).

Finally, conclusions and implications of our results appear in
Section~\ref{sec:conc}, and some technical details and supporting
calculations appear in the appendices.

\section{``Dynamic'' vortex mass}
\label{sec:dynamic}

This section takes a slight detour. We will review earlier
computations on the influence of phase fluctuations on the vortex
mass. Consideration of the influence of nodal quasiparticles will
begin in Section~\ref{secVortexAction}.

Popov\cite{popov} and Duan\cite{duan} have argued that vortices
acquire a ``dynamic'' mass from their coupling to quantum
fluctuations of the phase of the superconducting order, and that
this contribution diverges logarithmically for a neutral
superfluid. We now present the analog of this effect in our
formalism, and show further that the divergence is absent in
charged superfluids.

We begin with a discussion of neutral superfluids. For this case,
in the action for the phase fluctuations in Eq.~(\ref{sa}), the
pre-factor $k$ in the last term is replaced by $k^2$: this ensures
only short-range interaction proportional to the square of the
local density fluctuation $(\nabla \times \vec{\mathcal{A}})/(2
\pi)$. After a suitable rescaling of coordinates to absorb factors
of the velocity of spin-waves, the resulting action can be written
in the compact `relativistic' form
\begin{equation}
\mathcal{S}_{\mathcal{A}n} = \frac{c}{2}\int \frac{d^3 p}{8 \pi^3}
 \mathcal{A}_\mu \left( \delta_{\mu\nu} p^2 - p_\mu p_\nu
\right) \mathcal{A}_\nu, \label{san}
\end{equation}
where $c$ is a coupling constant and $p_\mu = (\omega, \Bf{k})$ is
a spacetime 3-momentum. Now we can use $\mathcal{S}_v$ in
Eq.~(\ref{sv}) and integrate out the $\mathcal{A}_\mu$
fluctuations under the action $\mathcal{S}_{\mathcal{A}n}$. In
this manner, we obtain the following additional term in the vortex
effective action
\begin{eqnarray}
\mathcal{S}_{v\mathcal{A}} &=& \frac{1}{2c} \int d \tau d \tau'
\int \frac{d^2 k d
\omega}{8 \pi^3} \frac{e^{-i \omega( \tau - \tau')}}{(\omega^2 + k^2)} \nonumber \\
&\times&  \left( 1 + \frac{d \Bf{r}_{\Tr{v}} (\tau)}{d \tau} \frac{d
\Bf{r}_{\Tr{v}} (\tau')}{d \tau'} \right) e^{i \Bf{k} \cdot (
\Bf{r}_{\Tr{v}} (\tau) - \Bf{r}_{\Tr{v}} (\tau'))} \label{sva1}
\end{eqnarray}
Integrating over $k$ and $\omega$, we obtain for large
$\tau-\tau'$
\begin{eqnarray}
\mathcal{S}_{v\mathcal{A}} &\sim&  \int d \tau d \tau'  \left( 1 +
\frac{d \Bf{r}_{\Tr{v}} (\tau)}{d \tau} \frac{d \Bf{r}_{\Tr{v}}
(\tau')}{d
\tau'} \right) \nonumber \\
&~&~\times \frac{1}{\left[( \Bf{r}_{\Tr{v}} (\tau) - \Bf{r}_{\Tr{v}}
(\tau'))^2 + ( \tau - \tau')^2\right]^{1/2}}
\end{eqnarray}
As we will repeatedly do in all the computations of this paper, we
expand this effective action for $\Bf{r}(\tau)$ in powers of a
slowly varying $\Bf{r}_{\Tr{v}} (\omega)$. This yields one term with
the structure
\begin{equation}
\mathcal{S}_{v\mathcal{A}} \sim  \int d \omega \omega^2
|\Bf{r}_{\Tr{v}} ( \omega) |^2 \int_{\Lambda^{-1}}^\infty d \tau
\frac{\cos (\omega \tau )}{\tau} \label{sva3}
\end{equation}
(where $\Lambda$ is a high energy cutoff), and another with the
structure
\begin{equation}
\mathcal{S}_{v\mathcal{A}} \sim \int d \omega  |\Bf{r}_{\Tr{v}} (
\omega) |^2 \int_{\Lambda^{-1}}^\infty d \tau \frac{(1-\cos (\omega
\tau ))}{\tau^3}. \label{sva4}
\end{equation}
Both terms yield a contribution to $F_{\parallel} (\omega)$ in
Eq.~(\ref{tsv}) which is $\sim \omega^2 \ln (\Lambda/|\omega|)$.
This logarithm is an indication of the divergence of the dynamic
mass, as found earlier in Refs.~\onlinecite{popov,duan}.

Let us now extend the above analysis to the charged case of
interest to the cuprates, where the $\mathcal{A}_\mu$ action is in
Eq.~(\ref{sa}). Now it is convenient to work in the Coulomb gauge,
when the only potentially singular contribution to the vortex
action comes from the fluctuations of the spatial components
$\vec{\mathcal{A}}$. In this case, Eq.~(\ref{sva1}) is replaced by
\begin{eqnarray}
&& \mathcal{S}_{v\mathcal{A}} = \frac{\pi}{e^{\ast 2}} \int d \tau
d \tau' \int \frac{d^2 k d
\omega}{8 \pi^3} \frac{e^{-i \omega( \tau - \tau')}}{(k + \omega^2 /(2 \pi \rho_s e^{\ast 2}))} \nonumber \\
&&\times  \frac{d r_{vi} (\tau)}{d \tau} \left(\delta_{ij} -
\frac{k_i k_j}{k^2}  \right) \frac{d r_{vj} (\tau')}{d \tau'} e^{i
\Bf{k} \cdot ( \Bf{r}_{\Tr{v}} (\tau) - \Bf{r}_{\Tr{v}} (\tau'))}
\nonumber
\\
\label{sva2}
\end{eqnarray}
As above, we expand in powers of $\Bf{r}_{\Tr{v}} (\omega)$, and
perform the momentum and frequency space integrals to obtain an
expression like that in Eq.~(\ref{sva3}), but with the $1/\tau$
under the $\tau$ integral replaced by $1/\tau^3$. This $\tau$
integral is convergent in the infrared, and hence we obtain a
$\omega^2$ term in $F_{\parallel} (\omega)$  with a finite
co-efficient {\em i.e.\/} a {\em finite\/} renormalization of the
vortex mass.

\section{Effective vortex action}
\label{secVortexAction}

We begin by deriving a general effective action for a single
vortex.\cite{Simanek,AoZhu2,AoZhu,Han} We will focus on the
contributions of the fermionic quasiparticles, with the aim of
renormalizing the action $\mathcal{S}_v + \mathcal{S}_\mathcal{A}$
presented in Section~\ref{sec:intro}. The vortex will be treated
as a singular configuration of external gauge and supercurrent
fields in the quasiparticle Hamiltonian. We write a path-integral
for quasiparticles and immediately integrate them out, which gives
us the vortex action as a functional of the time-dependent vortex
location. This is done in an expansion in the vortex velocity.
Formally, for this procedure to be well defined, quasiparticles
must be massive. Therefore, we will initially give a finite mass
to the nodal quasiparticles, and send it to zero later when it
becomes safe.

The vortex action is the logarithm of the fermion determinant in
the path-integral:
\begin{equation}\label{VAct1}
S_{\Tr{v}} = - \Tr{tr} \log G^{-1} \ .
\end{equation}
Here, the time-ordered quasiparticle Green's function in imaginary time satisfies:
\begin{equation}\label{GreenEq}
\left( \frac{\partial}{\partial \tau} + H \right) G(\Bf{r},\tau;\Bf{r}',\tau') =
  \delta(\tau-\tau') \delta^2(\Bf{r}-\Bf{r}') I \ ,
\end{equation}
where $I$ is the unit-matrix for spin degrees of freedom. In a
$d$-wave superconductor there are four flavors of nodal
quasiparticles - their contributions must be calculated separately
and added to the vortex action.

The quasiparticle Hamiltonian $H$, which we discuss in the next
section, depends on the vortex position $\Bf{r}_{\Tr{v}}(\tau)$.
We expand it to the second order in small vortex displacements
from the origin:
\begin{eqnarray}
H & = & H_0 + (\Bf{r}_{\Tr{v}} \nabla_{\Tr{v}}) H_0 +
  \frac{1}{2} (\Bf{r}_{\Tr{v}} \nabla_{\Tr{v}})^2 H_0 + \cdots \\
  & = & H_0 + \Delta H(\Bf{r}_{\Tr{v}}) \ , \nonumber
\end{eqnarray}
where $\nabla_{\Tr{v}} = \frac{\partial}{\partial
\Bf{r}_{\Tr{v}}}$. We also expand the Green's function, and write
in the operator notation:
\begin{equation}
G^{-1} = G^{-1}_0 + \Delta H \ ,
\end{equation}
with $G_0$ being the unperturbed Green's function (vortex sitting
at the origin). Now expand ~(\ref{VAct1}):
\begin{equation}\label{VAct2}
S_{\Tr{v}} = - \Tr{tr} \log G^{-1}_0 - \Tr{tr} (G_0 \Delta H)
  + \frac{1}{2} \Tr{tr} (G_0 \Delta H)^2 + \cdots \ .
\end{equation}
The first term is a constant and we drop it. In order to evaluate
the other terms, we substitute the solution of ~(\ref{GreenEq})
for the unperturbed Green's function:
\begin{equation}\label{Green0}
G_0(\Bf{r},\tau;\Bf{r}',\tau') = \sum_{n} \frac{1}{\beta} \sum_{\omega}
  \frac{e^{-i\omega(\tau-\tau')}}{-i\omega + \epsilon_n}
  \psi_n^{\phantom{\dagger}}(\Bf{r}) \psi_n^{\dagger}(\Bf{r}') \ .
\end{equation}
The wavefunctions $\psi_n$ and energies $\epsilon_n$ must be
solutions of the Schr\"odinger equation in position
representation:
\begin{equation}\label{Schr}
H_0(\Bf{r}) \psi_n(\Bf{r}) = \epsilon_n \psi_n(\Bf{r}) \ ,
\end{equation}
and frequencies $\omega$ take discrete values $(2k+1)\pi T$ at
finite temperature $T = 1 / \beta$. Clearly, the vortex dynamics
can be derived entirely from the knowledge of quasiparticle
energies and wavefunctions in presence of the vortex at rest.

The second term of ~(\ref{VAct2}) expanded to the second order in vortex position is:
\begin{equation}
-\Tr{tr}(G_0 \Delta H) =
  -\Tr{tr} \left\lbrack G_0 (\Bf{r}_{\Tr{v}} \nabla_{\Tr{v}}) H_0 \right\rbrack
  -\frac{1}{2} \Tr{tr} \left\lbrack G_0 (\Bf{r}_{\Tr{v}} \nabla_{\Tr{v}})^2 H_0
  \right\rbrack \ . \nonumber
\end{equation}
In absence of disorder, translational symmetry requires that there
be no linear terms in $\Bf{r}_{\Tr{v}}(\tau)$ in the vortex action.
This is explicitly demonstrated in Appendix \ref{appVortexAction}.
At the second order in $\Bf{r}_{\Tr{v}}$ we have:
\begin{eqnarray}
& & -\frac{1}{2} \Tr{tr} \left\lbrack G_0 (\Bf{r}_{\Tr{v}} \nabla_{\Tr{v}})^2 H_0 \right\rbrack = \\
& & \quad = -\int \dd\tau\dd^2 r \ \Tr{tr} \left\lbrack G_0(\Bf{r},\tau;\Bf{r},\tau)
  (\Bf{r}_{\Tr{v}}(\tau) \nabla_{\Tr{v}})^2 H_0(\Bf{r}) \right\rbrack \ . \nonumber
\end{eqnarray}
Since there is only one time variable in this integral, it
produces a ``trapping'' potential $\sim
|\Bf{r}_{\Tr{v}}(\tau)|^2$, which seems to trap the vortex at the
origin even in absence of disorder. This term will, therefore, be
cancelled in the complete vortex action, as we will see below.

We are left with the third term of ~(\ref{VAct2}):
\begin{eqnarray}\label{Trace3a}
& &  \frac{1}{2} \Tr{tr} (G_0 \Delta H)^2 = \\
& &  \int \dd\tau \dd^2 r \int \dd\tau' \dd^2 r'
  \Tr{tr} \Bigl\lbrack G_0(\Bf{r},\tau;\Bf{r}',\tau')
  \left( \Bf{r}_{\Tr{v}}(\tau') \nabla_{\Tr{v}} H_0(\Bf{r}') \right) \nonumber \\
& & \qquad  G_0(\Bf{r'},\tau';\Bf{r},\tau)
  \left( \Bf{r}_{\Tr{v}}(\tau) \nabla_{\Tr{v}} H_0(\Bf{r}) \right) \Bigr\rbrack \ . \nonumber
\end{eqnarray}
Let us utilize the absence of disorder and introduce a convenient
notation for the following matrix elements:
\begin{eqnarray}\label{UMatrEl}
\Bf{U}_{n,n'}
  & = & - \int \dd^2 r \psi_n^{\dagger}(\Bf{r}) \Bf{\nabla}_{\Tr{v}} \psi_{n'}^{\phantom{\dagger}}(\Bf{r}) \\
  & = & \int \dd^2 r \psi_n^{\dagger}(\Bf{r}) \Bf{\nabla} \psi_{n'}^{\phantom{\dagger}}(\Bf{r})  \ . \nonumber
\end{eqnarray}
If we substitute ~(\ref{Green0}) into the trace ~(\ref{Trace3a}),
integrate out the time variables and partially sum over frequencies
(details of which are in Appendix \ref{appVortexAction}), we obtain
a simple expression:
\begin{eqnarray}\label{Trace3b}
& & \frac{1}{2} \Tr{tr} (G_0 \Delta H)^2 = \\
& & \frac{1}{2} \sum_{n,n'} \frac{1}{\beta} \sum_{\omega}
  \frac{f(\epsilon_n)-f(\epsilon_{n'})}
  {\epsilon_n-\epsilon_{n'}-i\omega}  (\epsilon_n - \epsilon_{n'})^2
  | \Bf{r}_{\Tr{v}}(\omega) \Bf{U}_{n,n'} |^2 \ , \nonumber
\end{eqnarray}
where $f(\epsilon) = 1/(1+e^{\beta\epsilon})$ is the Fermi-Dirac
distribution function. Note that in the last expression $\omega$
already describes vortex motion - it is a ``bosonic'' Matsubara
frequency, taking values $2k \pi T$ at finite temperatures. This
expression is not zero at $\omega=0$, which implies its
non-invariance under translations, even in absence of disorder.
However, the trapping potential encountered earlier can now be
exactly cancelled. A simple recipe, which is rigorously derived in
Appendix \ref{appVortexAction}, is to subtract from
~(\ref{Trace3b}) its zero-frequency part. This way we obtain the
complete effective vortex action:
\begin{eqnarray}\label{VActFull}
S_{\Tr{v}} & = & \frac{1}{2} \sum_{n,n'} \frac{1}{\beta} \sum_{\omega}
  \left( f(\epsilon_n) - f(\epsilon_{n'}) \right)
  \frac{i\omega(\epsilon_n - \epsilon_{n'})}{\epsilon_n-\epsilon_{n'}-i\omega} \times \nonumber \\
& &  | \Bf{r}_{\Tr{v}}(\omega) \Bf{U}_{n,n'} |^2  \ .
\end{eqnarray}

The vortex action must be invariant under all lattice symmetries of
the superconducting material. Translational symmetry is already
guaranteed, while the $90^o$ degree rotation symmetry of the square
lattice is implemented only after the contributions from all
quasiparticle nodes are included. Symmetry under $90^o$ rotations
generates the same requirements on the vortex action at the second
order in vortex displacement as the symmetry under any rotation. The
allowed rotationally invariant terms are therefore
$|\Bf{r}_{\Tr{v}}(\omega)|^2$ and
$i\hat{\Bf{z}}(\Bf{r}_{\Tr{v}}^*(\omega) \times
(\Bf{r}_{\Tr{v}}(\omega))$, and we may write the action as:
\begin{equation}\label{VActLT}
S_{\Tr{v}} = \frac{1}{\beta} \sum_{\omega} \Bigl\lbrack
  F_{\parallel}(\omega) |\Bf{r}_{\Tr{v}}(\omega)|^2 +
  F_{\perp}(\omega) i\hat{\Bf{z}}(\Bf{r}_{\Tr{v}}^*(\omega) \times (\Bf{r}_{\Tr{v}}(\omega))
  \Bigr\rbrack \ .
\end{equation}
The first term describes vortex mass and dissipation, while the
second term breaks the time-reversal symmetry and describes possible
Magnus-type forces on the vortex arising from presence of
quasiparticles.

\section{Quasiparticle Hamiltonian}\label{secHamiltonian}

In this section we derive a convenient linearized form of the
quasiparticle Hamiltonian that will be used to find the effective
vortex action. While our focus will be the Bogoliubov-de Gennes
(BdG) description of quasiparticles, we note that similar results
can also be obtained from the ``staggered-flux'' model of
quasiparticles in $d$-wave superconductors.

The BdG Hamiltonian for an extreme type-II $d$-wave superconductor
is:
\begin{equation}\label{FullBdG}
H_{\Tr{BdG}} = \left(
  \begin{array}{cc}
    \frac{(\Bf{p}-\frac{e}{c}\Bf{A})^2}{2m} - E_F & \Delta  \\
    \Delta^* & -\frac{(\Bf{p}+\frac{e}{c}\Bf{A})^2}{2m} + E_F
  \end{array}
\right) \ ,
\end{equation}
where $\Bf{p}$ is the momentum operator, $\Bf{A}$ gauge field,
$E_F = \hbar^2k_F^2 / 2m$ quasiparticle Fermi
energy, and $\Delta$ gap operator given by:
\begin{eqnarray}
\Delta & = & \frac{1}{\hbar^2k_F^2} \lbrace p_x , \lbrace p_y , \Delta(\Bf{r})
  \rbrace \rbrace - \frac{i}{4k_F^2} \Delta(r) \lbrace \partial_x,\partial_y \rbrace
  \Phi(\Bf{r}) \nonumber \\
& = & \frac{\Delta(\Bf{r})}{\hbar^2k_F^2}
  \left\lbrace p_x + \frac{\hbar \partial_x \Phi(\Bf{r})}{2} ,
          p_y + \frac{\hbar \partial_y \Phi(\Bf{r})}{2} \right\rbrace \ .
\end{eqnarray}
$\Delta(\Bf{r})$ is the superconducting center-of-mass complex gap
function, with phase $\Phi(\Bf{r})$, and amplitude $\Delta_0$
assumed to be uniform (except at vortex cores). The braces denote
anticommutators defined as $\lbrace a,b \rbrace =
\frac{1}{2}(ab+ba)$. This Hamiltonian formally has $d_{xy}$
symmetry, which is related to $d_{x^2-y^2}$ by a rotation.

We follow the approach from Ref.~\onlinecite{Tesh1}
and~\onlinecite{Tesh2}. It was developed for a vortex lattice, but
we will use it in the limit of infinite inter-vortex separation. A
vortex lattice can be partitioned into two sublattices, $\Tr{A}$
and $\Tr{B}$. Let $\Phi_{\Tr{A}}$ and $\Phi_{\Tr{B}}$ be
superfluid phases that originate only from the vortices on
$\Tr{A}$ and $\Tr{B}$ sublattices respectively, so that
$\Phi(\Bf{r}) = \Phi_{\Tr{A}}(\Bf{r}) + \Phi_{\Tr{B}}(\Bf{r})$.
Define:
\begin{eqnarray}\label{GFdef}
m\Bf{v}_{\Tr{A}}(\Bf{r}) & = & \left( \hbar \nabla\Phi_{\Tr{A}}(\Bf{r}) -
  \frac{e}{c}\Bf{A}(\Bf{r}) \right) \nonumber \\
m\Bf{v}_{\Tr{B}}(\Bf{r}) & = & \left( \hbar \nabla\Phi_{\Tr{B}}(\Bf{r}) -
  \frac{e}{c}\Bf{A}(\Bf{r}) \right) \\
\Bf{a}(\Bf{r}) & = & \frac{m\Bf{v}_{\Tr{A}}(\Bf{r})-m\Bf{v}_{\Tr{B}}(\Bf{r})}{2} \nonumber \\
\Bf{v}(\Bf{r}) & = & \frac{\Bf{v}_{\Tr{A}}(\Bf{r})+\Bf{v}_{\Tr{B}}(\Bf{r})}{2}
  \ . \nonumber
\end{eqnarray}
In the extreme type-II limit the London penetration depth is
large, and the magnetic field may appear practically uniform
across inter-vortex distances. In such circumstances one can
neglect contribution of the gauge field to the total flux of
$m\Bf{v}$ on $\Tr{A}$ and $\Tr{B}$ sublattices and keep only the
$2\pi$ contribution of the superfluid order parameter. Then, the
fields $\Bf{a}(\Bf{r})$ and $\Bf{v}(\Bf{r})$ look like vector
potentials of a $\pi$-flux vortex lattice, where the flux alters
sign between $\Tr{A}$ and $\Tr{B}$ sublattices in
$\Bf{a}(\Bf{r})$, but not in $\Bf{v}(\Bf{r})$, which is the
superfluid velocity.

Now perform a unitary Franz-Te\v sanovi\'c transformation
\cite{ft} $U$ on the Hamiltonian ~(\ref{FullBdG}):
\begin{equation}
U = \left(
  \begin{array}{cc}
    e^{-i\Phi_{\Tr{A}}(\Bf{r})} & 0  \\
    0 & e^{i\Phi_{\Tr{B}}(\Bf{r})}
  \end{array}
\right) \ ,
\end{equation}
and obtain:
\begin{equation}
H_{\textrm{BdG}} = \left(
  \begin{array}{cc}
    \frac{(\Bf{p}+m\Bf{v}_{\Tr{A}})^2}{2m} - E_F & D  \\
    D & -\frac{(\Bf{p}-m\Bf{v}_{\Tr{B}})^2}{2m} + E_F
  \end{array}
\right) \ ,
\end{equation}
with the transformed gap operator:
\begin{equation}
D = \frac{\Delta_0}{\hbar^2 k_F^2}
  \lbrace p_x + a_x , p_y + a_y \rbrace \ .
\end{equation}
Finally, this Hamiltonian can be linearized near any of the four nodal points:
\begin{eqnarray}\label{BdG}
    & & \Tr{near} \pm \hbar k_F \hat{\Bf{x}} \ : \\[1mm]
      H & = & \pm \left\lbrack v_F (p_x + a_x) \sigma^z +
      v_{\Delta} (p_y + a_y) \sigma^x + m v_F v_x I \right\rbrack \nonumber \\[2mm]
    & & \Tr{near} \pm \hbar k_F \hat{\Bf{y}} \ : \nonumber \\[1mm]
      H & = & \pm \left\lbrack v_F (p_y + a_y) \sigma^z +
      v_{\Delta} (p_x + a_x) \sigma^x + m v_F v_y I \right\rbrack \ , \nonumber
\end{eqnarray}
where $\sigma^x, \sigma^y, \sigma^z$ are Pauli matrices, and $I$
is the unit two-by-two matrix. The last term in these
Hamiltonians, proportional to $I$, is the Doppler shift of
quasiparticle energy due to the background superfluid flow.

Before we take the limit of infinite inter-vortex separation, it
is important to understand how the magnetic field varies with
distance from a vortex core. A more accurate expression for the
superfluid velocity $\Bf{v}(\Bf{r})$ is obtained by solving the
conventional London equation (Appendix of
Ref.~\onlinecite{Tesh2}):
\begin{equation}\label{SFvel}
\Bf{v}(\Bf{r}) = \frac{2\pi\hbar}{m} \int\frac{\dd^2 k}{(2\pi)^2}
  \frac{i\Bf{k} \times \hat{\Bf{z}}}{k^2} \left( 1 - \frac{1}{1+\lambda^2 k^2} \right)
  C_{\Bf{k}} e^{i \Bf{kr}} \ ,
\end{equation}
where $\lambda$ is the London penetration depth, and:
\begin{equation}
C_{\Bf{k}} = \sum_i \frac{e^{-i\Bf{kr}_i^{\Tr{A}}} + e^{-i\Bf{kr}_i^{\Tr{B}}}}{2} \ .
\end{equation}
Note that the field $\Bf{a}(\Bf{r})$ does not depend on $\lambda$;
all of $\lambda$ dependence is contained in the Doppler shift of
~(\ref{BdG}). The superfluid velocity $\Bf{v}(\Bf{r})$ decays as
$1/r$ with distance from the vortex core when $\lambda\to\infty$,
and faster for finite $\lambda$.

\section{Analytic results in absence of the Doppler shift}\label{secAnalytic}

The expression ~(\ref{VActFull}) for the effective vortex action,
and the description of quasiparticles given by ~(\ref{BdG}) provide
us with all ingredients we need to study vortex dynamics.

We will follow two routes to the study of this dynamics, and the
reader may examine them in either order.

A perturbative computation is presented in
Section~\ref{sec:perturbation}, and the reader can jump ahead to
that section now. There we begin with an alternative
formulation\cite{sf,bf,grsf,bs} of the same theory using auxiliary
Chern-Simons gauge fields. This  allows a straightforward
consideration of the structure of the (uncontrolled) perturbation
theory, which leads to the results in Eq.~(\ref{nonohmic}).

In the second route, we use a non-perturbative analysis, and show
that certain physical effects can be evaluated exactly. This route
proceeds in two stages. First, we will ignore the Doppler shift of
quasiparticle energies in the present section, and this will allow
us to analytically derive the vortex action. Such analytic results
will elucidate interesting physics, and serve as a framework for the
second stage (section \ref{secNumerics}) in which we include the
Doppler shift. Calculations of the Doppler shift effects must be
done numerically, but fortunately, no qualitative modifications of
the analytic partial results are found. All these results are also
consistent with Eq.~(\ref{nonohmic}).

\subsection{Transition matrix elements}\label{ssecMatrEl}

The main problem is to diagonalize the linearized BdG Hamiltonian
~(\ref{BdG}) and calculate the matrix elements $\Bf{U}_{n,n'}$ given
by  ~(\ref{UMatrEl}), which need to be substituted into the vortex
action ~(\ref{VActFull}). The matrix elements $\Bf{U}_{n,n'}$ can be
evaluated exactly in absence of the Doppler shift. For now we focus
on the node $\Bf{k} = \hbar k_F\hat{\Bf{x}}$, and postpone
discussion of the other nodes for later. We start by solving the
Schr\"odinger equation ~(\ref{Schr}) for the Hamiltonian:
\begin{equation}\label{HNoDoppler}
H = \left(
  \begin{array}{cc}
    v_F(p_x+a_x) & v_{\Delta}(p_y+a_y)  \\
    v_{\Delta}(p_y+a_y) & -v_F(p_x+a_x)
  \end{array}
\right) \ .
\end{equation}
We will approximate the ``gauge'' field $\Bf{a}(\Bf{r})$ to that
of a $\pi$-flux vortex completely localized at the origin. Even
though the vortex core will appear infinitely small, the
approximation is validated by the fact that there are no strictly
bound quasiparticle states in a $d$-wave vortex, so that the
vortex dynamics is determined only by extended states.

We will determine the spectrum of $H$ following the analysis of
Ref.~\onlinecite{Tesh2}. As they showed, microscopic details of
the vortex core can be captured in a single parameter. The
Hamiltonian ~(\ref{HNoDoppler}) could be easily diagonalized if it
were isotropic. Therefore, we make it isotropic by applying a
gauge transformation that deforms the vortex gauge field
$\Bf{a}(\Bf{r})$ to a suitable ``elliptical'' shape, and then
rescale the coordinates in such a way that $v_F$ and $v_{\Delta}$
become equal to unity and the ``elliptical'' gauge field is
reshaped to the isotropic form:
\begin{equation}\label{VGF}
\Bf{a}(\Bf{r}) = \frac{-y\hat{\Bf{x}}+x\hat{\Bf{y}}}{2(x^2+y^2)} \ .
\end{equation}
The coordinate rescaling goes as follows (the original coordinates are ``primed''):
\begin{equation}
  \begin{array}{lcl}
     w_x = v_F w'_x , w_y = v_{\Delta} w'_y & \dots & \Tr{vectors} \\[3mm]
     \dd^2 r' = v_F v_{\Delta} \dd^2 r      & \dots & \Tr{Jacobian} \\[3mm]
     \nabla' = \frac{\hat{\Bf{x}}}{v_F}\frac{\partial}{\partial_{x}}
            + \frac{\hat{\Bf{y}}}{v_{\Delta}}\frac{\partial}{\partial_{y}}
            & \dots & \Tr{gradient} \\[3mm]
     \psi'(x',y') = \frac{1}{\sqrt{v_F v_{\Delta}}} \psi(\frac{x}{v_F},\frac{y}{v_{\Delta}})
            & \dots & \Tr{wavefunctions} \\[3mm]
     H = (p_x+a_x)\sigma^z + (p_y+a_y)\sigma^x & \dots & \Tr{Hamiltonian}
  \end{array} \nonumber
\end{equation}
It is convenient to apply a unitary transformation to the rescaled Hamiltonian and write it as:
\begin{equation}
H = \left(
  \begin{array}{cc}
    -m & (p_x+a_x)-i(p_y+a_y)  \\
    (p_x+a_x)+i(p_y+a_y) & m
  \end{array}
\right) \ ,
\end{equation}
where a quasiparticle mass $m$ has been added ``by hand''. In a
$d$-wave superconductor $m=0$, and eventually we set $m$ to zero,
but formally a non-zero mass is needed to integrate out
quasiparticles in the path-integral and perform summations over
frequencies such as ~(\ref{OmegaSum1}). Switching to the polar
coordinates:
\begin{equation}\label{DerPol}
  \begin{array}{lcl}
    p_x = \cos\phi \left( -i\frac{\partial}{\partial_r} \right)
          -\frac{\sin\phi}{r} \left( -i\frac{\partial}{\partial_{\phi}} \right)
    & \quad & a_x = -\frac{\sin\phi}{2r} \\
    p_y = \sin\phi \left( -i\frac{\partial}{\partial_r} \right)
          +\frac{\cos\phi}{r} \left( -i\frac{\partial}{\partial_{\phi}} \right)
    & \quad & a_y = \frac{\cos\phi}{2r}
  \end{array}
\end{equation}
the Schr\"odinger equation takes form:
\begin{widetext}
\begin{equation}
\left(
  \begin{array}{cc}
    -m & e^{-i\phi} \left( -i\frac{\partial}{\partial_r} - \frac{i}{r}
       \left( -i\frac{\partial}{\partial_{\phi}} + \frac{1}{2} \right) \right) \\
     e^{i\phi} \left( -i\frac{\partial}{\partial_r} + \frac{i}{r}
       \left( -i\frac{\partial}{\partial_{\phi}} + \frac{1}{2} \right) \right) & m
  \end{array}
\right)
\left(
  \begin{array}{c}
     u(\Bf{r}) \\
     v(\Bf{r})
  \end{array}
\right)
= E
\left(
  \begin{array}{c}
     u(\Bf{r}) \\
     v(\Bf{r})
  \end{array}
\right)
\ .
\end{equation}
\end{widetext}
By substituting $u(\Bf{r}) = e^{i(l-1)\phi} u(r)$ and $v(\Bf{r}) = e^{il\phi} v(r)$ the solutions are found to be:
\begin{eqnarray}\label{WF}
& & \psi_{q,l,k}(r,\phi) = \left( \frac{k}{4\pi|E|} \right)^{\frac{1}{2}} \times \\
& &
\left\lbrace
  \begin{array}{lcl}
    \left(
      \begin{array}{c}
        \sqrt{E-m} \cdot J_{-l+\frac{1}{2}}(kr) e^{i(l-1)\phi} \\
        -iq\sqrt{E+m} \cdot J_{-l-\frac{1}{2}}(kr) e^{il\phi}
      \end{array}
    \right)
    & \quad , \quad & l < 0 \\[5mm]
    \left(
      \begin{array}{c}
        \sqrt{E-m} \cdot J_{l-\frac{1}{2}}(kr) e^{i(l-1)\phi} \\
        iq\sqrt{E+m} \cdot J_{l+\frac{1}{2}}(kr) e^{il\phi}
      \end{array}
    \right)
    & \quad , \quad & l > 0
  \end{array}
\right\rbrace \ . \nonumber
\end{eqnarray}
States are characterized by the quantum numbers $n=(q,l,k)$:
``charge'' $q = \pm 1$ (distinguishes particles-like and hole-like
states), angular momentum $l \in \mathbb{Z}$ and radial wavevector
$k>0$. Energy is $E = q \sqrt{k^2+m^2}$. $J_l(kr)$ are Bessel
functions of the first kind. These wavefunctions are normalized as:
\begin{equation}
\int \dd^2 r \psi^{\dagger}_{q_1,l_1,k_1}(\Bf{r})
  \psi^{\phantom{dagger}}_{q_2,l_2,k_2}(\Bf{r}) =
  \delta_{q_1,q_2} \delta_{l_1,l_2} \delta(k_1-k_2) \ . \nonumber
\end{equation}
The zero angular momentum channel requires special attention. Square integrability of the wavefunctions requires only that:
\begin{eqnarray}\label{WF0}
& & \psi_{q,0,k}(r,\phi) =  \left( \frac{k}{4\pi|E|} \right)^{\frac{1}{2}} \times \\
& &
  \begin{array}{lcl}
    \Biggl\lbrack
      \sin\theta
      \left(
        \begin{array}{c}
          \sqrt{E-m} \cdot J_{-\frac{1}{2}}(kr) e^{-i\phi} \\
          iq\sqrt{E+m} \cdot J_{\frac{1}{2}}(kr)
        \end{array}
      \right) + & & \\[5mm]
      \cos\theta
      \left(
        \begin{array}{c}
          \sqrt{E-m} \cdot J_{\frac{1}{2}}(kr) e^{-i\phi} \\
          -iq\sqrt{E+m} \cdot J_{-\frac{1}{2}}(kr)
        \end{array}
      \right)\Biggr\rbrack
      & \quad \dots \quad & l=0
  \end{array}
\nonumber
\end{eqnarray}
where $\theta$ is a parameter with arbitrary value within
$[0,\pi)$. This parameter is physically obtained from boundary
conditions at the vortex core only when the core is not idealized
by a point, but regarded as a finite region in space. Microscopic
details of the vortex core enter vortex dynamics only through this
parameter. In the following we assume that all of the flux is
contained in a small cylindrical shell: this yields $\theta=0$ for
a vortex, and $\theta=\frac{\pi}{2}$ for an anti-vortex
\cite{Tesh1}. Other choices do not introduce qualitative changes
in the results that follow, but they may give rise to some
quantitative corrections, to the vortex mass for example, which we
briefly discuss in the Appendix \ref{appTheta}.

The presence of a vortex significantly affects quantum motion of
quasiparticles, as is clearly seen from the local density of states
(LDOS) expressed in the rescaled coordinates (for $m=0$):
\begin{eqnarray}\label{LDOS}
\rho(\epsilon; r,\theta) & = &
   \frac{\cos(2|\epsilon|r)}{2 \pi^2 r} + \frac{|\epsilon|}{\pi} \sum_{l=0}^{\infty}
   J_{l+\frac{1}{2}}^2 (|\epsilon|r) \nonumber \\
& \longrightarrow & \left\lbrace
       \begin{array}{c@{\quad , \quad}c}
          \frac{1}{2 \pi^2 r} & |\epsilon|r \ll 1 \\[2mm]
          \frac{|\epsilon|}{2\pi} & |\epsilon|r \gg 1
       \end{array} \right\rbrace  \ .
\end{eqnarray}
At small distances from the vortex core LDOS diverges as $1/r$ due
to the zero angular momentum state ~(\ref{WF0}), and there are
also certain oscillations. This behavior was studied earlier, and
with addition of the Doppler shift it only acquires anisotropy,
but does not qualitatively deviate from $1/r$ dependence
\cite{Melnikov}. In realistic circumstances this divergence is
cut-off by the finite core size. Note that the LDOS peak at $r=0$
does not correspond to bound states in purely $d$-wave vortex
cores (bound states could emerge only if a secondary order
parameter, such as s-wave, were present in vicinity of the
vortex). In fact, a careful analysis shows that the Hamiltonian
~(\ref{HNoDoppler}) does not support even a bound state exactly at
$E=0$ \cite{ZeroBoundStates}. Far away from the vortex core LDOS
becomes constant and linear in energy as expected.

We can now calculate the matrix elements $\Bf{U}_{n,n'}$ in
~(\ref{UMatrEl}). This calculation can be done directly in the
rescaled coordinate system, with the states ~(\ref{WF}) and
~(\ref{WF0}) that were obtained after a gauge transformation,
rescaling and a unitary transformation:
\begin{equation}\label{U2}
\Bf{U}_{n_1,n_2} = \int\dd^2 r \psi^{\dagger}_{n_1}(\Bf{r})
  \left( \frac{\hat{\Bf{x}}}{v_F}\frac{\partial}{\partial_{x}}
       + \frac{\hat{\Bf{y}}}{v_{\Delta}}\frac{\partial}{\partial_{y}} \right)
  \psi^{\phantom{\dagger}}_{n_2}(\Bf{r}) \ .
\end{equation}
Here $n_i$ denotes $(q_i,l_i,k_i)$.

The calculation of ~(\ref{U2}) is pretty tedious, so that we only
sketch it in the Appendix \ref{appUMatrEl}. A closed form for
arbitrary quasiparticle mass $m$ is obtained by algebraic
manipulations and elementary integrations, with help of Bessel
function identities. It turns out that no singularities are
introduced in the end by taking the $m \to 0$ limit, so that in
the following we will consider only the case of interest $m=0$.
Since $\Bf{U}_{n_1,n_2}$ are essentially transition matrix
elements of the momentum operator, the only allowed transitions
are between states whose angular momentum is different by one.
Defining $\sigma = l_2 - l_1 = \pm 1$, we obtain for the node
$\hbar k_F \hat{\Bf{x}}$ and zero quasiparticle mass:
\begin{equation}\label{UExpr}
\Bf{U}_{n_1, n_2} = \frac{1}{8} e^{\frac{i\pi}{4}(q_2-q_1)}
  \left( \sigma \frac{\hat{\Bf{x}}}{v_F} + i \frac{\hat{\Bf{y}}}{v_{\Delta}} \right)
  U_{(q_1,l-\sigma,k_1),(q_2,l,k_2)} \ ,
\end{equation}
where\begin{widetext}
\begin{equation}\label{UScalar}
U_{(q_1,l-\sigma,k_1),(q_2,l,k_2)} = \left\lbrace
  \begin{array}{lcl}
    4 \sqrt{k_1 k_2} \delta(E_2 - E_1) -
      C_{\sigma; q_1, q_2} \left( \frac{k_1}{k_2} \right)^{\sigma l - \frac{1}{2}}
      \frac{E_1 + E_2}{\sqrt{k_1 k_2}}
      \Theta \left( \sigma (k_2 - k_1) \right)
    & \quad , \quad & l > \frac{\sigma + 1}{2} \\
    - 4 \sqrt{k_1 k_2} \delta(E_1 - E_2) -
      C_{\sigma; q_1, q_2} \left( \frac{k_1}{k_2} \right)^{\sigma l - \frac{1}{2}}
      \frac{E_1 + E_2}{\sqrt{k_1 k_2}}
      \Theta \left( \sigma (k_1 - k_2) \right)
    & \quad , \quad & l < \frac{\sigma + 1}{2} \\
    \frac{2 \sigma q_1 q_2}{\pi}\frac{E_1+E_2}{E_1-E_2} +
      \frac{1}{2\pi}
      C_{\sigma; q_1, q_2} \left( \frac{k_1}{k_2} \right)^{\sigma l - \frac{1}{2}}
      \frac{E_1 + E_2}{\sqrt{k_1 k_2}}
      \log \left( \frac{k_1-k_2}{k_1+k_2} \right)^2
    & \quad , \quad & l = \frac{\sigma + 1}{2}
  \end{array}
\right\rbrace \ .
\end{equation}
\end{widetext}
Here, $\delta(x)$ is the Dirac delta-function, and:
\begin{equation}
\Theta(x) = \left\lbrace
  \begin{array}{lcl}
    1 & \quad , \quad & x>0 \\
    \frac{1}{2} & \quad , \quad & x=0 \\
    0 & \quad , \quad & x<0
  \end{array}
\right\rbrace \ ; \nonumber
\end{equation}
\begin{equation}
C_{\sigma; q_1, q_2} = \left\lbrace
  \begin{array}{lcl}
     q_2 & \quad , \quad & \sigma=1 \\
     q_1 & \quad , \quad & \sigma=-1
  \end{array}
\right\rbrace \ . \nonumber
\end{equation}
Quasiparticle energies are now $E_i = q_i k_i$. The expression
~(\ref{UScalar}) defines infinities at $E_1=E_2$ completely, except
when $l=(\sigma+1)/2$. In that case, however, it can be shown by
finite-size regularization that $U$ at  $E_1=E_2$ diverges only
logarithmically with the system size, which does not affect the
important integrals over energies in the limit of infinite system
size beyond what is given by the expression ~(\ref{UScalar}).

\subsection{Summation over quasiparticle nodes and angular momenta}

The effective vortex action ~(\ref{VActFull}) depends on $\Bf{U}_{n_1,n_2}$ through:
\begin{eqnarray}\label{U2Expr0}
& & | \Bf{r}_{\Tr{v}}(\omega) \Bf{U}_{n_1,n_2} |^2 =
  \left\vert \frac{U_{(q_1,l-\sigma,k_1),(q_2,l,k_2)}}{8} \right\vert^2 \times \\
& & \hspace*{-4.5mm} \left\lbrack
  \left( \frac{|\hat{\Bf{x}}\Bf{r}_{\Tr{v}}(\omega)|^2}{v_F^2} +
         \frac{|\hat{\Bf{y}}\Bf{r}_{\Tr{v}}(\omega)|^2}{v_{\Delta}^2} \right) +
  \frac{i\sigma}{v_F v_{\Delta}} \hat{\Bf{z}} \left(
         \Bf{r}^*_{\Tr{v}}(\omega) \times \Bf{r}_{\Tr{v}}(\omega) \right) \right\rbrack  , \nonumber
\end{eqnarray}
where we have used ~(\ref{UExpr}), relevant for the quasiparticle
node at $\Bf{p} = \hbar k_F \hat{\Bf{x}}$. Note that this dependence
on vortex coordinates is not isotropic; only final expressions with
all quasiparticle nodes included become isotropic. It is easily
checked from ~(\ref{BdG}) that the linearized Hamiltonians at
different nodes are related by unitary transformations and an
exchange $v_F \leftrightarrow v_{\Delta}$ as needed, in absence of
the Doppler shift. Thus, in order to add contributions at all nodes
we only need to symmetrize with respect to $v_F$ and $v_{\Delta}$
and multiply by the number of nodes:
\begin{eqnarray}\label{U2Expr}
& & \hspace{-4mm} \sum_{\Tr{nodes}} | \Bf{r}_{\Tr{v}}(\omega) \Bf{U}_{n_1,n_2} |^2 =
   \frac{|U_{(q_1,l-\sigma,k_1),(q_2,l,k_2)}|^2}{32} \times \\
& &
  \left\lbrack \left( \frac{1}{v_F^2} + \frac{1}{v_{\Delta}^2} \right) |\Bf{r}_{\Tr{v}}(\omega)|^2 +
  \frac{2i\sigma}{v_F v_{\Delta}} \hat{\Bf{z}} \left(
         \Bf{r}^*_{\Tr{v}}(\omega) \times \Bf{r}_{\Tr{v}}(\omega) \right) \right\rbrack \ . \nonumber
\end{eqnarray}
Note that for our purposes we do not need to worry about inter-node
scattering of quasiparticles. If we went back to the non-linear
quasiparticle Hamiltonian and calculated $\Bf{U}_{n_1,n_2}$ for
states that belong to different nodes, we could obtain singular
behavior only when the momentum transfer is $\sim \hbar k_F$. Such
transitions are not stimulated by a slowly moving vortex, and cannot
play a significant role in low-energy vortex dynamics.

Owing to the degeneracy of the quasiparticle energies with respect
to angular momentum, it is possible to analytically carry out
summation over angular momentum channels in ~(\ref{VActFull}), that
is summation over $l$ and $\sigma$. Vortex mass and possible
dissipation are determined by the following sum, which is just a
geometric series that we evaluate from ~(\ref{UScalar}):
\begin{eqnarray}\label{USumL}
& & \hspace{-4mm} \sum_{l,\sigma} \left\vert U_{(q_1,l-\sigma,k_1),(q_2,l,k_2)} \right\vert^2 =
  \frac{(E_1+E_2)^2 (E_1^2+E_2^2)}
       {|E_1^2-E_2^2| \Tr{max}(E_1^2,E_2^2)} \nonumber \\
& + & \left( \frac{E_1+E_2}{\pi} \right)^2 \Biggl\lbrack
    \frac{8}{(E_1-E_2)^2} +
    \frac{2}{k_1 k_2} \log \left( \frac{k_1-k_2}{k_1+k_2} \right)^2 \nonumber \\
& + &  \frac{1}{4} \left( \frac{1}{E_1^2}+\frac{1}{E_2^2} \right)
    \log^2 \left( \frac{k_1-k_2}{k_1+k_2} \right)^2
  \Biggr\rbrack  \ .
\end{eqnarray}
Here, we have dropped the Dirac delta-functions in ~(\ref{UScalar}),
since in ~(\ref{VActFull}) there is a factor that becomes zero when
$E_1 = E_2$ (safe to do only in absence of the Doppler shift). For
the same reason the other possible divergences at $k_1 = k_2$, in
~(\ref{UScalar}) or in the sum over angular momenta turn out not be
dangerous: they actually appear only when $q_1 = q_2$ as well, that
is only when $E_1 = E_2$ (particle-particle or hole-hole
transitions).

Similarly, we find the sum that determines ``transversal dynamics'':
\begin{eqnarray}\label{USumT}
& & \hspace{-4mm} \sum_{l,\sigma} \sigma \left\vert U_{(q_1,l-\sigma,k_1),(q_2,l,k_2)} \right\vert^2 = \nonumber \\
& & \frac{(E_1+E_2)^2}{\Tr{max}(E_1^2, E_2^2)} \Tr{sign}(|E_1|-|E_2|) \nonumber \\
& & \hspace{-3mm} + \left( \frac{E_1+E_2}{\pi} \right)^2 \Biggl\lbrack
    \frac{2 q_1 q_2}{E_1 - E_2} \left( \frac{1}{E_1} + \frac{1}{E_2} \right)
    \log \left( \frac{k_1-k_2}{k_1+k_2} \right)^2 \nonumber \\
& & \hspace{-3mm} + \frac{1}{4} \left( \frac{1}{E_2^2} - \frac{1}{E_1^2} \right)
    \log^2 \left( \frac{k_1-k_2}{k_1+k_2} \right)^2
  \Biggr\rbrack  \ .
\end{eqnarray}

\subsection{Vortex mass at zero temperature}

This subsection collects our results above and computes the
contribution of the $\pi$-flux to the vortex mass. We will find that
this contribution is finite.

The vortex mass is determined by the ``longitudinal'' part of the
vortex action ~(\ref{VActLT}). Thus we will restrict our attention
to the sum ~(\ref{USumL}) and substitute it into the vortex action
~(\ref{VActFull}). At zero temperature the difference between
Fermi-Dirac functions $f(\epsilon_{n_1}) - f(\epsilon_{n_2})$ allows
only transitions between states whose energies have opposite signs.
This means that only particle-hole transitions will be allowed.
Using the invariance of ~(\ref{USumL}) under exchange $E_1
\leftrightarrow E_2$, and symmetrizing the ``longitudinal'' part of
~(\ref{VActFull}) with respect to its complex conjugate, we have:
\begin{eqnarray}\label{VActT0}
S_{\Tr{v} \parallel} & = & \int\frac{\dd\omega}{2\pi}
  \int\limits_{-\Lambda}^{0}\dd E_1 \int\limits_{0}^{\Lambda}\dd E_2
  \frac{\omega^2 (E_2 - E_1)}{(E_2 - E_1)^2 + \omega^2} \times \nonumber  \\
& &  \sum_{\Tr{nodes}}\sum_{l,\sigma} | \Bf{r}_{\Tr{v}}(\omega) \Bf{U}_{n_1,n_2} |^2
\ .
\end{eqnarray}
Energy cut-off is $\Lambda$. If slow vortex dynamics reduces to
vortex inertia, then the leading low-frequency term in the
``longitudinal'' vortex action is:
\begin{equation}
S_{\Tr{v} \parallel} = \int\frac{\dd\omega}{2\pi}
\frac{m_{\Tr{v}}\omega^2}{2}
  |\Bf{r}_{\Bf{\Tr{v}}}(\omega)|^2 \ ,
\end{equation}
where $m_{\Tr{v}}$ is the vortex mass:
\begin{eqnarray}\label{VortexMass}
m_{\Tr{v}} & = & \frac{1}{16} \left( \frac{1}{v_F^2} +
\frac{1}{v_{\Delta}^2} \right)
  \int\limits_{-\Lambda}^{0}\dd E_1 \int\limits_{0}^{\Lambda}\dd E_2 \\
& & \frac{1}{E_2-E_1}
  \sum_{l,\sigma}
  \left\vert U_{(q_1,l-\sigma,k_1),(q_2,l,k_2)} \right\vert^2 \ . \nonumber
\end{eqnarray}
Note that this expression is obtained by simply setting $\omega=0$
in the denominator of ~(\ref{VActT0}), which is justified only if
$m_{\Tr{v}}$ turns out to be finite. In absence of thermal
fluctuations $m_{\Tr{v}}$ could in principle be infra-red divergent
only due to transitions at $E_1 = E_2 = 0$. However, only the bulk
density of states, which shapes such transitions, can be responsible
for infra-red divergences. In the bulk, far away from the vortex
core, LDOS is linear in energy (see ~(\ref{LDOS})) and hence
suppresses transitions at $E=0$. This is formally reflected in the
sum ~(\ref{USumL}) by absence of the singularity at $E_1=0^+$,
$E_2=0^-$. Then, without help from the transition matrix elements,
the extra factor of $E_2-E_1$ in ~(\ref{VortexMass}) is innocuous
and fully integrable. Note that the $1/r$ behavior of LDOS close to
the vortex core matters, but not enough to cause infra-red
divergency of $m_{\Tr{v}}$; it produces the second term in
~(\ref{USumL}), including the logarithms, but no singularity at
$E=0$.


Careful integration, presented in Appendix \ref{appVortexMass}, shows that indeed the vortex mass is finite:
\begin{equation}\label{VMassNoDoppler}
m_{\Tr{v}}  \approx 0.05 \left( \frac{1}{v_F^2} +
\frac{1}{v_{\Delta}^2} \right) \Lambda \ .
\end{equation}
Therefore, vortex dynamics at zero temperature is entirely specified by
inertia. The cut-off scale $\Lambda$ is roughly given by the
maximum superconducting gap $\Delta_0$. We can write:
\begin{equation}\label{VMassNoDoppler2}
m_{\Tr{v}}  \sim 0.05 \left( \alpha_D + \frac{1}{\alpha_D} \right)
m_e \ ,
\end{equation}
where $m_e$ is the electron mass, and $\alpha_D = v_F/v_\Delta$. In
$d$-wave superconductors $\alpha_D \sim 1 \div 10^2$, so that the
purely quantum effects contribute roughly an electron mass to the
mass of a vortex.

In various circumstances, which are not the focus of this paper, the
bulk LDOS may not vanish at zero energy. For example, this can be
expected in presence of disorder, or Zeeman splitting in magnetic
field. In such situations the vortex mass calculated from the
expression ~(\ref{VortexMass}) could become infra-red divergent, and
it would be more appropriate to go back to the full action
~(\ref{VActT0}).

\subsection{Dissipation at finite temperatures}\label{ssecFiniteT}

At finite temperatures the ``longitudinal'' part of the vortex action is:
\begin{eqnarray}\label{VActT1}
S_{\Tr{v}\parallel} & = & \frac{1}{2\beta} \sum_{\omega}
  \int\limits_{-\Lambda}^{\Lambda}\dd E_1 \int\limits_{-\Lambda}^{\Lambda}\dd E_2
  \left( f(E_1) - f(E_2) \right) \times \nonumber \\
& &  \hspace{-3mm} \frac{\omega^2 (E_2 - E_1)}{(E_2 - E_1)^2 + \omega^2}
        \sum_{\Tr{nodes}}\sum_{l,\sigma} | \Bf{r}_{\Tr{v}}(\omega) \Bf{U}_{n_1,n_2} |^2 \ .
\end{eqnarray}
The difference between Fermi-Dirac functions $f(E_1) - f(E_2)$ does
not strictly prohibit transitions between two particle-like or two
hole-like states at $T>0$. Therefore, singularities in which the
factors $E_2-E_1$ become zero at finite energies will be included in
the vortex action. We can no longer naively neglect $\omega^2$ in
the denominator, even for small frequencies.

Consider the ``longitudinal correlation'' $F_{\parallel}(\omega)$
defined by ~(\ref{VActLT}). At small frequencies we expect that
small values of $E_2-E_1$ will matter the most, so that we
approximate:
\begin{equation}
f(E_1) - f(E_2) \approx (E_1 - E_2)
  \frac{\partial f}{\partial E}
    \Bigl\vert_{\frac{E_1+E_2}{2}}
  \quad , \quad |E_1-E_2| \ll T \ . \nonumber
\end{equation}
With a change of variables $\Delta E = E_2-E_1$, $2E = E_2 + E_1$ we have at small frequencies:
\begin{eqnarray}
& & \hspace{-6mm} F_{\parallel}(\omega) \approx \frac{1}{64}  \left( \frac{1}{v_F^2} + \frac{1}{v_{\Delta}^2} \right)
   \int\limits_{-T}^{T}\dd \Delta E \int\limits_{-\Lambda}^{\Lambda}\dd E
    \left( -\frac{\partial f(E)}{\partial E} \right) \times \nonumber \\
& &  \hspace{-2mm} \frac{\omega^2 \Delta E^2}{\Delta E^2 + \omega^2}
        \sum_{l,\sigma} | U_{(q_1,l-\sigma,k_1),(q_2,l,k_2)} |^2 \ .
\end{eqnarray}
We do not care about $|\Delta E| > T$, because for sufficiently large and finite $\Delta E$ we can safely expand in powers of $\omega^2$ and obtain only the high-frequency corrections to vortex inertia. Our goal here is to elucidate sub-quadratic powers of frequency. Similarly, it will be clear shortly that the precise boundaries of $\Delta E$ do not matter after analytic continuation. The sum over angular momenta is given by ~(\ref{USumL}). Its ``most divergent'' part ($2^{\Tr{nd}}$ term), which comes from the particle-particle and hole-hole transitions that involve the zero angular momentum channel, dominates the small frequency behavior:
\begin{eqnarray}\label{FPT}
F_{\parallel}(\omega) & \approx & \frac{1}{64}  \left( \frac{1}{v_F^2} + \frac{1}{v_{\Delta}^2} \right)
   \int\limits_{-T}^{T}\dd \Delta E \int\limits_{-\Lambda}^{\Lambda}\dd E
    \left( -\frac{\partial f(E)}{\partial E} \right) \times \nonumber \\
& &  \hspace{-2mm} \frac{\omega^2 \Delta E^2}{\Delta E^2 + \omega^2}
       \left( \frac{32}{\pi^2} \frac{E^2}{\Delta E^2} + \cdots \right) \\
& \approx & \frac{\pi}{6} \left( \frac{1}{v_F^2} +
\frac{1}{v_{\Delta}^2} \right) T^2 |\omega|
     + \cdots
    \ . \nonumber
\end{eqnarray}
After analytic continuation to real frequency, this has the
``Ohmic'' dissipation form of the Bardeen-Stephen viscous drag. The
ellipses denote corrections that become more noticeable as the
frequency grows. The first such correction is also interesting and
comes from particle-particle and hole-hole transitions that do not
involve the zero angular momentum channel ($1^{\Tr{st}}$ term of
~(\ref{USumL})):
\begin{eqnarray}
\Delta F_{\parallel}(\omega) & \approx & \frac{1}{64}  \left( \frac{1}{v_F^2} + \frac{1}{v_{\Delta}^2} \right)
   \int\limits_{-T}^{T}\dd \Delta E \int\limits_{-\Lambda}^{\Lambda}\dd E
    \left( -\frac{\partial f(E)}{\partial E} \right) \nonumber \\
& \times & \frac{\omega^2 \Delta E^2}{\Delta E^2 + \omega^2}
  \frac{4E^2 (2E^2 + \frac{\Delta E^2}{2})}{|2E \Delta E| (E^2 + \frac{\Delta E^2}{4} + |E \Delta E|)} \nonumber \\[2mm]
& & \hspace{-5mm} \approx \frac{\log(2)}{8}  \left( \frac{1}{v_F^2} + \frac{1}{v_{\Delta}^2} \right)
   T \omega^2 \log \left( 1 + \frac{T^2}{\omega^2} \right) + \cdots
    . \nonumber
\end{eqnarray}
This could be interpreted as a frequency-dependent correction to
the vortex mass. With such an interpretation the vortex mass at rest would be
logarithmically divergent. However, we feel that it is more natural to
associate only the strictly quadratic frequency behavior to the
vortex mass, and regard everything that dominates it at small
frequencies as various forms of dissipation. All other corrections
are $\mathcal{O}(\omega^2)$, and hence contribute the normal
vortex mass.

In conclusion, dissipation occurs at finite temperatures, its
strength being a power law in temperature due to the gapless
quasiparticle nodes.

\subsection{Transversal dynamics}\label{ssecTransDyn}

The ``transversal correlation'' $F_{\perp}(\omega)$ defined in
~(\ref{VActLT}) describes aspects of vortex dynamics that are
associated with the Magnus force. We argue below that the nodal
quasiparticles do not give rise to transversal forces on a vortex at
any temperature.

The expression for ``transversal correlations'' is:
\begin{eqnarray}\label{TransF}
F_{\perp}(\omega) & = & \frac{1}{32 v_F v_{\Delta}}
  \int\limits_{-\Lambda}^{\Lambda}\dd E_1 \int\limits_{-\Lambda}^{\Lambda}\dd E_2
  \left( f(E_1) - f(E_2) \right) \times \nonumber \\
& & \hspace{-8mm} \frac{-i\omega (E_2 - E_1)}{E_2 - E_1 - i\omega}
       \sum_{l,\sigma} \sigma | U_{(q_1,l-\sigma,k_1),(q_2,l,k_2)} |^2 \ .
\end{eqnarray}
The sum over angular momenta is given by ~(\ref{USumT}); its
crucial property as a function of $E_1$ and $E_2$ is that it
changes sign under the change of variables $E_1 \to -E_2$ and $E_2
\to -E_1$. If we apply this change of variables to
$F_{\perp}(\omega)$, and add the result to the original expression
above, we obtain:
\begin{eqnarray}
& & \hspace{-4mm} 2F_{\perp}(\omega) \propto
  \int\limits_{-\Lambda}^{\Lambda}\dd E_1 \int\limits_{-\Lambda}^{\Lambda}\dd E_2
  \sum_{l,\sigma} \sigma | U_{(q_1,l-\sigma,k_1),(q_2,l,k_2)} |^2 \times \nonumber \\
& & \left( f(E_1) + f(-E_1) - f(E_2) - f(-E_2) \right) \frac{-i\omega (E_2 - E_1)}{E_2 - E_1 - i\omega} \nonumber \\[2mm]
& & = 0 \ . \nonumber
\end{eqnarray}
The final conclusion follows from the identity $f(E)+f(-E)=1$.

The physical reason for complete absence of quasiparticle-induced
transversal forces is related to the fact that there are no
circulating currents of quasiparticles in thermal equilibrium.
This connection has been made at small frequencies in
Ref.~\onlinecite{AoZhu}. At small frequencies it is safe to
neglect $\omega$ in the denominator of ~(\ref{TransF}), and then
cancel out the factors of $E_2-E_1$. What remains is an integral
that can be represented as a pure trace of a quantum operator, and
this trace can be evaluated in real space. Using the definition of
transition matrix elements ~(\ref{UMatrEl}), one finds:
\begin{equation}
F_{\perp}(\omega) \propto -i\omega \int \dd^2 r \Bf{\hat{z}} (\Bf{\nabla} \times \Bf{j}) = -i\omega \oint \Bf{j} \dd \Bf{s} \ ,
\end{equation}
where $\Bf{j}$ is the net current density of quasiparticles,
determined by wavefunctions ~(\ref{WF}), ~(\ref{WF0}), and
temperature. The contour integral above is taken on a loop of
infinite radius. In thermal equilibrium quasiparticles are at rest
with respect to the substrate, so that $\Bf{j}=0$.

\section{Perturbative computation}
\label{sec:perturbation}

This section reviews an alternative formulation\cite{sf,bf,grsf,bs}
of the quantum theory of vortices, fermionic quasiparticles, and
plasmons which appears in Sections~\ref{sec:intro}
and~\ref{secHamiltonian}. In this approach, the complete action for
the vortices, plasmons, and nodal quasiparticles is
$\widetilde{\mathcal{S}}_v + \mathcal{S}_\mathcal{A} +
\mathcal{S}_\Psi + \mathcal{S}_{cs}$ specified in Eqs.~(\ref{svf}),
(\ref{sa}), (\ref{spsif}), and (\ref{scs}). This theory is closely
related to a continuum limit of the theory considered by Balents and
Fisher.\cite{bf}

The new feature of this approach is that the Berry phase terms are
mediated by auxiliary Chern-Simons gauge fields. Such a formalism
readily allows a perturbative analysis of the couplings between the
elementary excitations using standard techniques. It must be kept in
mind, however, that there is no small parameter which justifies such
a perturbative analysis.

We will see below that the perturbative analysis of the Berry
phase terms lead to conclusions consistent with the exact
evaluation already presented in Section~\ref{secAnalytic}.
Empowered by this success, we will examine the influence of the
Doppler shift term using perturbation theory in
Section~\ref{secDopplerpert}.

We begin by extending the vortex action $\mathcal{S}_v$ in
Eq.~(\ref{sv}) by coupling it in addition to a second U(1) gauge
field $\alpha_\mu$:
\begin{eqnarray}
\widetilde{\mathcal{S}}_v &=& \int d\tau \Biggl[
\frac{m_{\Tr{v}}^0}{2} \left( \frac{d \Bf{r}_{\Tr{v}} ( \tau)}{d
\tau } \right)^2 + i \frac{d \Bf{r}_{\Tr{v}} ( \tau)}{d \tau } \cdot
\vec{\mathcal{A}}^0
(\Bf{r}_{\Tr{v}} ( \tau)) \nonumber \\
&+& i \int d^2 \Bf{r} d \tau \left\{ \mathcal{A}_{\mu} (\Bf{r},
\tau) + \alpha_\mu ( \Bf{r}, \tau ) \right\} J_{v\mu} (\Bf{r},
\tau) \Biggr] \label{svf}
\end{eqnarray}
This is to facilitate the Berry phase coupling to the fermionic
quasiparticles, as we now describe.

The fermionic quasiparticle Hamiltonian is discussed in
Sections~\ref{sec:intro} and~\ref{secHamiltonian}, and we work
with a single node appearing in the first equation in
Eq.~(\ref{BdG}). Let this Hamiltonian act on the Nambu spinor
$\Psi$. We define $\overline{\Psi} = - i  \Psi^{\dagger}
\sigma^y$, introduce the Dirac gamma matrices $\gamma^\mu =
(-\sigma^y, \sigma^x, \sigma^z)$, rescale coordinates and gauge
fields to obtain isotropic velocities as discussed in
Section~\ref{secAnalytic}, and thence obtain the continuum action
for the fermions
\begin{equation}
\mathcal{S}_\Psi = \int d^2 \Bf{r} d \tau \left[ - i
\overline{\Psi} \gamma^\mu ( \partial_\mu - i a_\mu ) \Psi +
\frac{iv_F}{2} \overline{\Psi} \gamma^0 \Psi \partial_x \Phi
\right].
\end{equation}
The last Doppler-shift term is a coupling to the superflow
fluctuations, which we have represented in this paper by the U(1)
gauge field $\mathcal{A}_\mu$. Extracting the explicit connection
between the gradient of the phase of the order parameter and the
`electric' field associated with $\mathcal{A}_\mu$, we may rewrite
this term to obtain the fermionic action in its final form
\begin{eqnarray}
\mathcal{S}_\Psi &=& \int d^2 \Bf{r} d \tau \Biggl[ - i
\overline{\Psi} \gamma^\mu ( \partial_\mu - i a_\mu ) \Psi
\nonumber
\\ &~&~~~~~~+ \frac{i v_F}{4 \pi \rho_s} \overline{\Psi} \gamma^0 \Psi
\left(\partial_y \mathcal{A}_\tau -
\partial_\tau \mathcal{A}_y \right) \Biggr]. \label{spsif}
\end{eqnarray}

Finally, we need to tie the gauge field $a_\mu$ to the vortices,
as was done explicitly in Eq.~(\ref{VGF}). In the present
formalism, this is done conveniently with a Chern-Simons term:
\begin{equation}
\mathcal{S}_{cs} = \frac{i}{\pi} \int d^2 \Bf{r} d \tau
\epsilon_{\mu\nu\lambda} a_\mu \partial_\nu \alpha_\lambda
\label{scs}
\end{equation}

We have now written all the required terms coupling together the
vortices, fermionic quasiparticles, and plasmons in a $d$-wave
superconductor. The complete action is $\widetilde{\mathcal{S}}_v
+ \mathcal{S}_\mathcal{A} + \mathcal{S}_\Psi + \mathcal{S}_{cs}$
specified in Eqs.~(\ref{svf}), (\ref{sa}), (\ref{spsif}), and
(\ref{scs}).

At this point, it is useful to describe how the results of
Sections~\ref{secHamiltonian}, \ref{secAnalytic}, and
\ref{secNumerics} relate to the present formalism. The Hamiltonian
in Eq.~(\ref{BdG}) is obtained by exactly integrating out the gauge
fields $a_\mu$, $\alpha_\mu$, and only the scalar potential
$\mathcal{A}_\tau$ in the Coulomb gauge $\nabla \cdot
\vec{\mathcal{A}} = 0$. The retardation effects contained in
$\vec{\mathcal{A}}$ fluctuations are neglected. The computations of
Section~\ref{secAnalytic} and \ref{secNumerics} describe the
consequences of subsequently exactly integrating out the fermions.

The present section proceeds by integrating out degrees of freedom
in the {\em opposite\/} order. We first integrate out the fermionic
fields $\Psi$ exactly, and then study the perturbative consequences
of subsequently integrating out the $a_\mu$, $\alpha_\mu$ and
$\mathcal{A}_\mu$. As we have already seen in
Section~\ref{sec:dynamic}, such an approach readily allows
accounting for the $\vec{\mathcal{A}}$ fluctuations.

We will first integrate out the fermions $\Psi$, and study its
influence on the $a_\mu$ and $\alpha_\mu$ gauge fields, and then
investigate the resulting modifications to the vortex/plasmon action
in Eqs.~(\ref{sv}) and (\ref{sa}). We defer consideration of the
Doppler shift term, and the resulting modification of plasmon ({\em
i.e.\/} $\mathcal{A}_\mu$) fluctuations to the following
Section~\ref{secDopplerpert}.

Integrating out $\Psi$ in Eq.~(\ref{spsif}) also yields an
effective action $\mbox{Tr} \ln [\gamma^\mu (\partial_\mu - i
a_\mu)]$ for $a_\mu$. This involves terms to all orders in $a_\mu$
and is not easy to work with. To obtain a first understanding of
its influence, we expand the functional determinant to second
order in $a_\mu$. Subsequently we integrate out the $a_\mu$, while
accounting for the Chern-Simons term in $\mathcal{S}_{cs}$. This
yields an action for the gauge field $\alpha_\mu$ of the form
\begin{equation}
\mathcal{S}_\alpha = \frac{\tilde{c}}{2}\int \frac{d^3 p}{8 \pi^3}
|p| \alpha_\mu \left( \delta_{\mu\nu} - \frac{p_\mu p_\nu}{p^2}
\right) \alpha_\nu, \label{salpha}
\end{equation}
where $\tilde{c}$ is a coupling constant. Now we can examine the
effect of $\alpha_\mu$ fluctuations on the vortices just as in
Section~\ref{sec:dynamic}. The action $\mathcal{S}_\alpha$ is
similar to that in Eq.~(\ref{san}) but with an extra power of $p$.
Consequently, we obtain results for the effective vortex action as
in Eqs.~(\ref{sva3}) and (\ref{sva4}) but with an extra power of
$1/\tau$ in the integrand. The $\tau$ integrals are now linearly
convergent in the infrared, and we obtain then a {\em finite\/}
renormalization of the vortex mass. This structure is entirely
consistent with the exact results of Section~\ref{secAnalytic}
where we found vortex mass contributions expressed in terms of
integrals which were also linearly convergent in the infrared.

More explicitly, in frequency and momentum space we obtain an
expression which is just as in Eq.~(\ref{sva1}), but with the
$(\omega^2 + k^2)$ in the denominator replaced by $\sqrt{\omega^2 +
k^2}$. Expanding the resulting expression in powers of
$\Bf{r}_{\Tr{v}} (\omega)$ we obtain a result in the form of
Eq.~(\ref{tsv}), with a contribution
\begin{equation}
F_{\parallel} (\omega ) = \frac{1}{2 \tilde{c}} \int \frac{d^2 k}{4
\pi^2} \frac{\left( k^2 - k \sqrt{k^2 + \omega^2} + 2 \omega^2
\right)}{2\sqrt{k^2 + \omega^2}}
\end{equation}
It is now easily seen that the result of the momentum integral (and
after reverting to physical units by re-inserting factors of the
velocities) has the form of Eq.~(\ref{nonohmic}), with $F_2$ a
universal function of the velocity ratio. In other words, we obtain
a finite non-universal renormalization of $m_{\Tr{v}}$, and a
universal ``sub-Ohmic'' dissipation.

\subsection{Doppler shift}
\label{secDopplerpert}

We are interested here in the influence of the last Doppler-shift
coupling in Eq.~(\ref{spsif}) between the nodal fermions and the
superflow (which is the dual `electric field'). At first glance,
it appears that this term is innocuous: integrating out the
fermions leads to an ultra-violet finite fermion loop, and this
contributes a term proportional to the square of the
$\mathcal{A}_\mu$ `electric' field with a finite co-efficient.
Comparing with Eq.~(\ref{sa}) we conclude that this is merely a
renormalization of the superfluid stiffness $\rho_s$ which
controls the plasmon fluctuations.

However, for completeness, we do have to also consider the
subleading non-analytic momentum and frequency dependence of the
fermion loop, which could lead to singular corrections to the vortex
action. We will see below that this does not happen, and that such
terms lead only to a finite renormalization of the vortex mass.

Let us write the electric field $\mathcal{E}_x = \partial_\tau
\mathcal{A}_x -
\partial_x \mathcal{A}_\tau$, and similarly for $\mathcal{E}_y$.
Then the term
\begin{equation}
\frac{1}{8 \pi^2 \rho_s} \int d^2 \Bf{r} \int d\tau \left(
\mathcal{E}_x^2 + \mathcal{E}_y^2 \right)
\end{equation}
in Eq.~(\ref{sa}) is replaced by
\begin{eqnarray}
&& \frac{1}{8 \pi^2 \rho_s} \int \frac{d^2 k}{4 \pi^2} \int
\frac{d \omega}{2 \pi} \left[ \left(1 + \frac{\mathcal{K} (k_x,
k_y, \omega)}{\rho_s} \right) |\mathcal{E}_y (\Bf{k}, \omega) |^2
\right.
\nonumber \\
&&~~~~~+ \left .\left( 1 + \frac{\mathcal{K} (k_y, k_x,
\omega)}{\rho_s} \right) |\mathcal{E}_x (\Bf{k}, \omega) |^2
\right]. \label{saf}
\end{eqnarray}
We compute the universal correction $\mathcal{K}$ at one loop
order, using the original unscaled units in the Hamiltonian in
Eq.~(\ref{BdG}). This leads to (after including contributions from
all nodes)
\begin{eqnarray}
&& \mathcal{K} (k_x, k_y, \omega) = \frac{v_F^2}{2} \int \frac{d^2
p}{4 \pi^2} \frac{d \epsilon}{2 \pi} \mbox{Tr} \nonumber
\\
&& \times \left[-i (\epsilon+\omega) + v_F (p_x + k_x) \sigma^z +
v_\Delta (p_y + k_y) \sigma^x \right]^{-1} \nonumber \\
&&\times \left[-i \epsilon + v_F p_x \sigma^z + v_\Delta p_y
\sigma^x \right]^{-1} \nonumber \\
&&~~~~~~~ =  v_F^2 \int \frac{d^2 p}{4 \pi^2} \frac{d \epsilon}{2
\pi} \left[ - \epsilon(\epsilon+\omega) + v_F^2 (p_x +k_x)p_x
\right. \nonumber \\ &&~~~~\left. + v_\Delta^2 (p_y + k_y) p_y
\right] \left[ \epsilon^2 + v_F^2 p_x^2
+ v_{\Delta} p_y^2 \right]^{-1} \nonumber \\
&&~~~~\times \left[ (\epsilon+\omega)^2 + v_F^2 (p_x + k_x)^2 +
v_{\Delta} (p_y + k_y)^2 \right]^{-1} \nonumber \\
&&~~~~~~~ = - \frac{ v_F(\omega^2 + 3 v_F^2 k_x^2 + 3 v_\Delta^2
k_y^2)}{64 v_\Delta \left[\omega^2 + v_F^2 k_x^2 + v_\Delta^2
k_y^2\right]^{1/2}}
\end{eqnarray}
where dimensional regularization was used in evaluating the
integrals (this is a rapid way of picking out the non-analytic
piece, as the analytic contribution has already been absorbed by
renormalizing $\rho_s$).

Now we can proceed to a computation of the renormalization of the
vortex action as in Section~\ref{sec:dynamic} except that the
substitution in Eq.~(\ref{saf}) has to be performed in the gauge
field action in Eq.~(\ref{sa}). As before, we work in the Coulomb
gauge. We focus on the $\mathcal{A}_\tau$ fluctuations, and the
expression in Eq.~(\ref{sva2}) is now replaced by
\begin{eqnarray}
&& \widetilde{\mathcal{S}}_{v\mathcal{A}} = 2 \pi^2 \rho_s \int d
\tau d \tau' \int \frac{d^2 k d
\omega}{8 \pi^3} e^{-i \omega( \tau - \tau')} \nonumber \\
&&\times \left[ k_y^2 \left(1 + \frac{\mathcal{K} (k_x, k_y,
\omega)}{\rho_s} \right)  + k_x^2 \left( 1 + \frac{\mathcal{K}
(k_y, k_x, \omega)}{\rho_s} \right) \right]^{-1} \nonumber \\
&&~~~~~~~~~~~~\times e^{i \Bf{k} \cdot ( \Bf{r}_{\Tr{v}} (\tau) -
\Bf{r}_{\Tr{v}} (\tau'))} \label{sva6}
\end{eqnarray}
Expanding this result as usual in small $\Bf{r}_{\Tr{v}} ( \omega)$,
and also to leading order in $\mathcal{K}$, we obtain the form in
Eq.~(\ref{tsv}) with
\begin{eqnarray}
F_{\parallel} ( \omega ) &=& - \pi^2 \int \frac{d^2 k}{(2 \pi)^2}
\frac{1}{k^2} \left[ k_y^2 ( \mathcal{K} (k_x, k_y, \omega) -
\mathcal{K} (k_x, k_y, 0)) \right. \nonumber \\
&+& \left. k_x^2 ( \mathcal{K} (k_y, k_x, \omega) - \mathcal{K}
(k_y, k_x, 0)) \right] \label{Gammaval}
\end{eqnarray}
It is interesting to note that all factors of $\rho_s$ have dropped
out of the expression for $F_{\parallel} (\omega)$, and a universal
expression dependent only upon the velocities $v_F$ and $v_\Delta$
has been obtained. The momentum integral in Eq.~(\ref{Gammaval}) is
also free of infrared divergences. However, an ultraviolet
divergence is present, and then the result has the form claimed
earlier in Eq.~(\ref{nonohmic}) with $F_2$ a universal function of
the velocity ratio.

It can be verified in a similar manner that the fluctuations of the
spatial component of the gauge field, $\vec{\mathcal{A}}$ lead to a
non-universal correction to the vortex mass.

\section{Influence of the Doppler shift: beyond perturbation theory}\label{secNumerics}

Section~\ref{secAnalytic} has provided insight into the vortex
dynamics with neglected Doppler shift. Then we included the effect
of the Doppler shift in Section~\ref{secDopplerpert}, but only in
perturbation theory. Now we go back to the full linearized
quasiparticle Hamiltonian ~(\ref{BdG}) and describe a computation
which includes the Doppler shift beyond perturbation theory. We will
only focus on an isolated vortex at zero temperature (some interesting effects in presence of other vortices are discussed in the Appendix \ref{ssecVInt}). In such
circumstances, only the ``longitudinal'' dynamics is non-trivial.
The main questions that we ask are how the vortex mass is modified,
and whether dissipation or some anomalous dynamics emerges at small
frequencies due to the Doppler shift. We will find that even after
accounting for the Doppler shift, the vortex mass remains finite, as
anticipated in the perturbation theory in the previous section.

It was hinted in Section~\ref{secAnalytic} that infra-red divergent
vortex mass, or various forms of dissipation, could emerge at zero
temperature if the bulk LDOS were finite at zero energy. Generally,
one would expect that the presence of a vortex does not alter the
bulk quasiparticle LDOS. This was found to be true in absence of the
Doppler shift, and should remain true when the Doppler shift is
included, since it is a localized perturbation (at any finite London
penetration depth). Then, the vanishing zero-energy bulk LDOS of an
ideal $d$-wave superconductor leads to a finite quasiparticle
contribution to the vortex mass. This scenario is intuitive, and
needs to be verified by a rigorous calculation. We will also verify
in this section whether the Doppler shift introduces dangerous
changes in LDOS near the vortex core. The concern about possible
anomalous vortex dynamics also comes  form a different point of
view. The anomalous dynamics was found in Section~\ref{secAnalytic}
whenever particle-particle or hole-hole transitions were allowed.
With the Doppler shift viewed as a perturbation, the exact
eigenstates will be superpositions of states ~(\ref{WF}) and
~(\ref{WF0}), so that the exact particle-like states will have
``hole tails''. Even the particle-hole transitions between exact
states, allowed at zero temperature, will incorporate hole-hole
transitions between the unperturbed states.


In the following we will exactly diagonalize the Hamiltonian
~(\ref{BdG}) on a sample with finite size $R$, and at the node
$\Bf{p} = \hbar k_F\Bf{\hat{x}}$; the other nodes are
easily included on symmetry grounds. Only the worst case scenario of infinite London penetration depth will be considered. We will use the states
~(\ref{WF}) and ~(\ref{WF0}) as the basis states for our
representation. The main reason for choosing these basis states is
that the transition matrix elements between them are analytically
known. This way all singularities, which are difficult to handle
numerically, will be captured inside analytical expressions.
Therefore, even a rough information about wavefunctions in the
thermodynamic limit $R \to \infty$, obtained from numerical finite
size scaling, might suffice to establish properties of the vortex
action.

At zero temperature we adapt the expression ~(\ref{VActFull}) and
write the longitudinal part of the vortex action:
\begin{eqnarray}\label{RFunct}
S_{\Tr{v} \parallel} & = & \int\frac{\dd\omega}{2\pi}
  \int\limits_{0}^{2\Lambda}\dd\Delta\epsilon
  \frac{\omega^2 \ \Delta\epsilon}{\Delta\epsilon^2 + \omega^2}
  \mathcal{R}(\Delta\epsilon) | \Bf{r}_{\Bf{\Tr{v}}}(\omega) |^2 \\
\mathcal{R}(\Delta\epsilon) & \propto &
  \sum_{n,n'}
    \Theta(\epsilon_n) \Theta(-\epsilon_{n'})
    \delta(\epsilon_n - \epsilon_{n'} - \Delta\epsilon)
    | \Bf{U}_{n,n'} |^2 \ . \nonumber
\end{eqnarray}
We will use labels $n$ for the exact eigenstates, and $\epsilon_n$
for exact energies. The basis states will be labeled by their
angular momentum $l$ and energy $E$, with $q=\Tr{sign}(E)$ and
$k=|E|$ understood. The exact transition matrix elements
$\Bf{U}_{n,n'}$, defined by ~(\ref{UMatrEl}), take the following
form in the chosen representation:
\begin{eqnarray}
\Bf{U}_{n_1,n_2} & = &
  \sum_{l_1,l_2} \int\limits_{-\Lambda}^{\Lambda} \dd E_1 \int\limits_{-\Lambda}^{\Lambda} \dd E_2
  \la E_1,l_1 | n_1 \ra \la n_2 | E_2,l_2 \ra \times \nonumber \\
& & \Bf{U}_{(E_1,l_1),(E_2,l_2)} \ ,
\end{eqnarray}
where the matrix elements $\Bf{U}_{(E_1,l_1),(E_2,l_2)}$ between
the basis states are given by ~(\ref{UExpr}).

\subsection{Exact quasiparticle spectrum and states}\label{ssecSpectrum}

We carried out numerical diagonalizations with typically $N_k = 50
\div 100$ discrete values for the radial wavevector $k$. This
merely sets the energy cut-off to $\Lambda \sim N_k R^{-1}$, and
provides a plenty of states with $kR \gg 1$ that we are interested
in. The number of angular momentum channels $l$ used in
calculations was varied between $N_l = 9 \div 29$. Ideally, $N_l
\sim \Lambda R$ is needed in order to include all possible states
with $|E|<\Lambda$ in the Hamiltonian representation, but this
would be too formidable. Fortunately, even though we worked with
small $N_l$, the final results showed virtually no dependence on
$N_l$, meaning that all interesting physics at low temperatures is
set by small angular momenta. The finite size scaling can be done
in two ways, by changing either $R$, or $\Lambda$. This is because
there is no specific scale in the linearized Hamiltonian and
thermodynamic limit (all coordinate dependence comes from the
products $kr$). Details of the Hamiltonian representation are shown in the Appendix \ref{appDoppler}.

The numerically obtained spectrum appears to be gapless. It also
appears symmetric under "particle-hole" exchange: for every state
with energy $\epsilon$ there is exactly one state with energy
$-\epsilon$. This property is, actually, protected by a symmetry of
the full Hamiltonian. There is a unitary transformation that
transforms the Hamiltonian $H$ in ~(\ref{BdG}) at any quasiparticle
node to $-H$, and this transformation is simply a coordinate system
rotation by $180^o$ degrees ($\phi \to \phi + \pi$ in ~(\ref{BdG})
after substitutions like ~(\ref{DerPol})). Therefore, the Doppler
shift does not cause ``spectral flow'', that is a shift of some
hole-like energy levels ($\epsilon<0$) into particle-like regions
($\epsilon>0$) or the other way round. The situation would have been
different if we considered a system of vortices, say a vortex
lattice, that is not inversion symmetric. If the $180^o$ rotation
symmetry were violated, ``spectral flow'' could happen. As analyzed
in Appendix~\ref{ssecVInt}, this could lead to dissipation and
anomalous dynamics at zero temperature.

\begin{figure}
\subfigure[{}]{\includegraphics[width=2.9in]{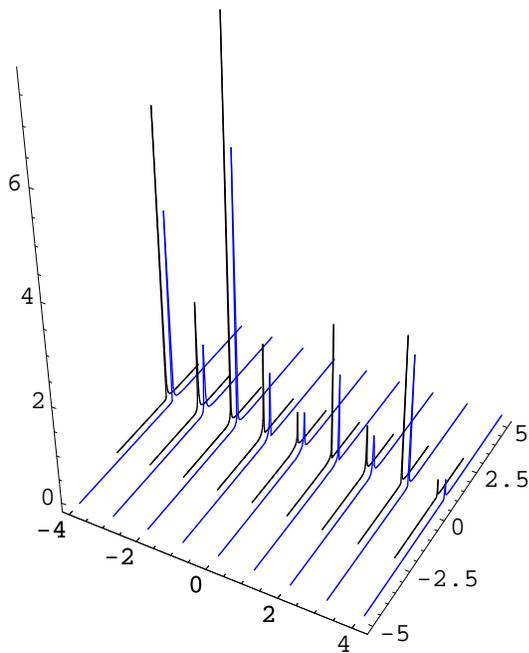}}
\subfigure[{}]{\includegraphics[width=2.9in]{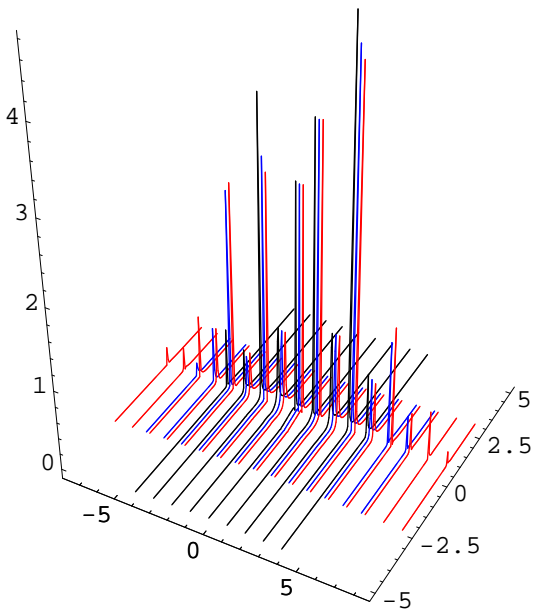}}
\caption{\label{wf-plots} (a) Typical wavefunction amplitude as a
function of energy and angular momentum for two system sizes, $R
\approx 63$ and $R \approx 125$. (b) Typical wavefunction amplitude
as a function of energy and angular momentum for three numbers of
angular momentum channels, $N_l=9$, $N_l=13$, and $N_l=17$.
Normalization is $\frac{R}{\pi}$, and $\alpha_D = 0.4$.}
\end{figure}

An example of numerically obtained eigenstates is plotted in
Figures~\ref{wf-plots}(a) and~\ref{wf-plots}(b). These figures show
the amplitude of $\Psi_n(E,l) = \la E,l | n \ra$ for a randomly
chosen eigenstate $|n\ra$ of the full Hamiltonian as a function of
$E$ and $l$ (which characterize the basis states - eigenstates in
absence of the Doppler shift). In order to gain insight about what
happens in the thermodynamic limit, the wavefunctions $\Psi_n(E,l)$
were normalized to the system radius $\frac{R}{\pi}$ instead of
usual unity. The Figure ~\ref{wf-plots}(a) compares two different
system sizes $R$. It can be noticed that roughly at $\epsilon \sim
E$ there is a single point where the amplitude is $|\Psi|^2 \propto
R$. This indicates that a Dirac peak may be developing in the
thermodynamic limit at $\epsilon=E$. At all other points, the
amplitudes $|\Psi|^2$ coincide for the different system sizes, and
there are ``tails'' of non-zero amplitude that spread away from
$\epsilon \sim E$, describable by finite analytic functions.
Possible survival of the Dirac peak can be understood in a
perturbative point of view where the strength of the Doppler shift
$H_D$ is parametrized and the full Hamiltonian written as $H = H_0 +
\lambda H_D$. For $\lambda=1$ we would obtain the Hamiltonian
~(\ref{BdG}) that is being numerically diagonalized. If $\lambda$
were zero, then the wavefunction $\Psi$ would be only the Dirac peak
by definition: $\Psi_{\epsilon,i}(E,l) = \delta_{i,l}
\delta(\epsilon-E)$, while for $\lambda>0$ the amplitude would start
leaking to other energies $\epsilon \neq E$, as was found
numerically. From the scattering theory we would expect that even
for $\lambda=1$ the Dirac peak survives at $\epsilon = E$; the
Doppler shift is a sufficiently localized perturbation to the
Hamiltonian that still has a singular gauge field.

\subsection{Vortex dynamics at small frequencies}

\begin{figure}
\includegraphics[height=2in]{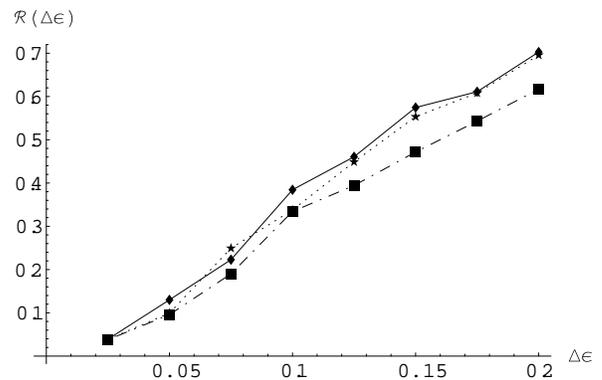}
\caption{\label{Rdiss} Function $\mathcal{R}(\Delta\epsilon)$ for
$\alpha_{\Tr{D}} = 10$, calculated with different numbers
$l_{\Tr{max}} \in \lbrace 4,6,8 \rbrace$ of included angular
momentum channels: $l \in \lbrace -l_{\Tr{max}}, \dots,
l_{\Tr{max}} \rbrace$. Both the horizontal and vertical axes are
expressed in units that scale as $1/R$ with system radius $R$.}
\end{figure}

\begin{figure}
\subfigure[{}]{\includegraphics[height=2in]{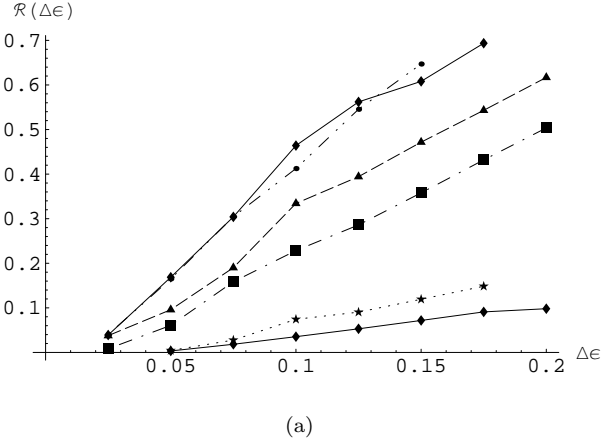}}
\subfigure[{}]{\includegraphics[height=2in]{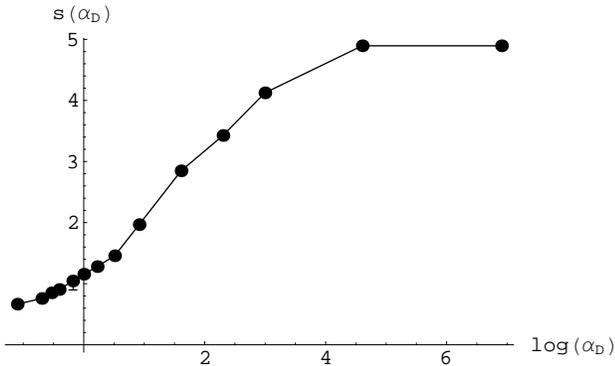}}
\caption{\label{Rcal} (a) Function $\mathcal{R}(\Delta\epsilon)$ for
$\alpha_{\Tr{D}} \in \lbrace 0.33, 1, 5, 10, 100, 1000 \rbrace$ (slope grows with $\alpha_{\Tr{D}}$), calculated with $l_{\Tr{max}} = 4$. Both the horizontal and vertical axes are expressed in units that scale as $1/R$ with system radius $R$. (b) Slope of $\mathcal{R}(\Delta\epsilon)$ (linear fit) as a function of the anisotropy ratio $\alpha_{\Tr{D}}$.}
\end{figure}

In order to obtain Ohmic dissipation, the function
$\mathcal{R}(\Delta\epsilon)$ in ~(\ref{RFunct}) should depend on
$\Delta\epsilon$ as $\Delta\epsilon^{-1}$. Actually,
$\Delta\epsilon^{\nu}$ with any $\nu \leq 0$ is a candidate for
anomalous behavior that dominates over finite vortex inertia at small
frequencies. The plot of $\mathcal{R}(\Delta\epsilon)$ evaluated
numerically is shown in the Figure \ref{Rdiss}. There are no
indications that any anomalous dynamics could occur. The situation
resembles very much the one that was encountered in absence of the
Doppler shift at zero temperature. The function
$\mathcal{R}(\Delta\epsilon)$ appears to be linear, so that its
slope $s$ from the Figure \ref{Rdiss} determines the vortex mass:
\begin{equation}
m_{\Tr{v}} = \frac{s(\alpha_D)}{8} \left( \alpha_{\Tr{D}} +
\frac{1}{\alpha_{\Tr{D}}} \right) m_e \ .
\end{equation}
For $\alpha_{\Tr{D}}=10$ this roughly equals $3.3$ electron masses,
which is a factor of $6.59$ enhancement of the vortex mass to that
obtained in absence of the Doppler shift. Clearly the Doppler shift
alone provides most of the vortex mass, but otherwise seems not to
alter the qualitative physics that was uncovered in Section
\ref{secAnalytic}. In Figure ~\ref{Rcal} we compare the functions
$\mathcal{R}(\Delta\epsilon)$ for several values of the anisotropy
factor $\alpha_{\Tr{D}}$, and plot the slope $s(\alpha_{\Tr{D}})$.
The slope $s(\alpha_D)$ does not change significantly above
$\alpha_D \ge 10$, so that in this range the vortex mass is roughly
proportional to the anisotropy factor $\alpha_D$. In the limit
$\alpha_{\Tr{D}} \to 0$ the matrix elements of the Doppler shift
vanish (see  Appendix \ref{appDoppler}), so that the slope
$s(\alpha_{\Tr{D}})$ slowly converges to $s \approx 8*0.05 = 0.4$
given by ~(\ref{VMassNoDoppler2}).

\begin{figure}
\subfigure[{}]{\includegraphics[height=2in]{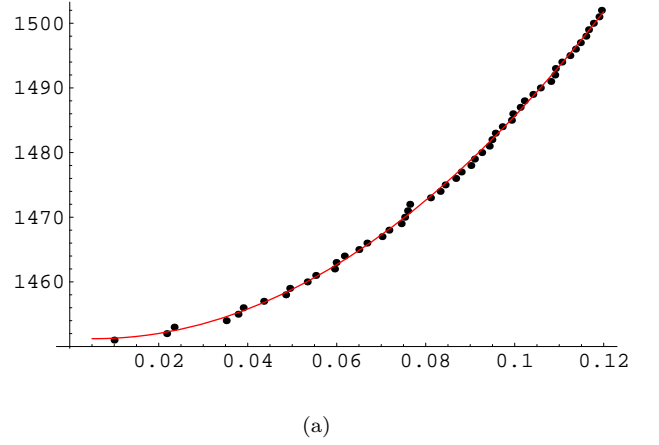}}
\subfigure[{}]{\includegraphics[height=2in]{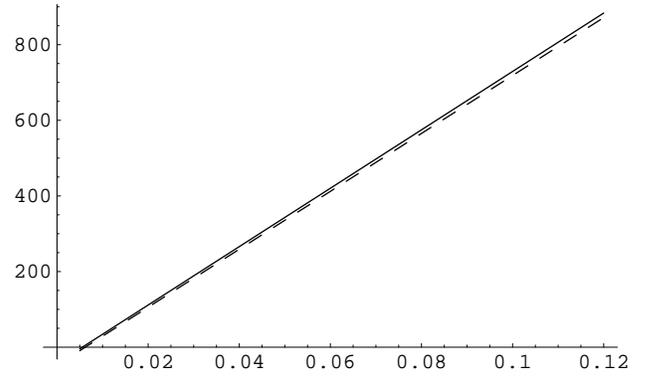}}
\caption{\label{DOS} (a) Total number of quasiparticle states in
presence of a vortex. Also shown is a quadratic fit to the data.
(b) Total density of states of the nodal quasiparticles in
presence of a vortex. Dashed line shows the density of states in
absence of the Doppler shift. Both plots have parameters
$\alpha_{\Tr{D}}=10$, $l_{\Tr{max}}=14$, and both the horizontal
and vertical axes are expressed in units that scale as $1/R$ with
system radius $R$.}
\end{figure}

Another indication that there should be no anomalous dynamics is
found in the quasiparticle density of states. The total density of
states at low energies in Figure \ref{DOS} is easily obtained from
the exact spectra, and it is indicative of the bulk LDOS. The bulk
LDOS remains a linear function of energy, and roughly equal to
LDOS in absence of the Doppler shift, as expected. Close to the
vortex core LDOS acquires anisotropy in the rescaled coordinates,
but does not deviate from the $1/r$ behavior. An asymptotic
solution \cite{Melnikov} for the wavefunctions that dominate LDOS
at $r \to 0$ in presence of the Doppler shift (at the node $\hbar
k_F \hat{\Bf{x}}$) is:
\begin{equation}
\psi(\Bf{r}) \propto \frac{1}{\sqrt{r}} \left(
  \begin{array}{l}
    c_1 \cos\left( \frac{\cos\phi}{2} \right) - c_2 \sin\left( \frac{\cos\phi}{2} \right) \\
    -e^{-i\phi} \left( c_1 \sin\left( \frac{\cos\phi}{2} \right) + c_2 \cos\left( \frac{\cos\phi}{2} \right) \right)
  \end{array}
\right) \ ,
\end{equation}
where $c_1$ and $c_2$ are constants fixed by matching to the regions of large $r$. The transition matrix elements ~(\ref{UMatrEl}) that involve these states and other normalizable states are always finite; the only possible concern are the ``diagonal'' matrix elements, but they must be finite as well because the states at finite energies carry finite momenta.


The last argument against dissipation and anomalous dynamics is
presented in the Appendix \ref{appSemiAnalytical}. There we study in
detail analytic properties of the exact transition matrix elements
$\Bf{U}_{n_1,n_2}$. Making use of the known transition matrix
elements ~(\ref{UExpr}) between the basis states in absence of the
Doppler shift, we ask the following general question: what should
the exact eigenstates look like in this basis in order for
dissipation to occur. The general conclusion is that dissipation
would require from the exact Hamiltonian to either systematically
shift or shuffle the energy levels of the basis states in vicinity
of the zero energy. The first scenario is equivalent to an effective
chemical potential for the linearized quasiparticles, and it was
discussed in the Appendix ~(\ref{ssecVInt}), while the second
scenario could happen due to Zeeman splitting or disorder. In both
cases the bulk LDOS would be finite. Neither of these scenarios seem
to be realized in our situation of interest, when we consider the
quantum-coherent Doppler shift due to the circulating supercurrents
of a single vortex.

\section{Discussion and conclusions}
\label{sec:conc}

We studied various effects introduced by nodal quasiparticles in clean
$d$-wave superconductors on dynamics of isolated vortices in
thermal equilibrium. These effects generally give dominant
contributions to vortex mass and dissipation in fermionic
superfluids, over the corresponding contributions of the
superfluid order parameter. In $d$-wave superconductors the
quasiparticles have gapless nodes, and this makes
quantum-mechanical phenomena fundamentally important for vortex
dynamics at low temperatures. Therefore, we used methods that
carefully took into account the phase change of quasiparticle
wavefunction upon encircling a vortex, resulting with the proper
determination of quasiparticle density of states. Classical
effects of the superfluid flow in the vicinity of a vortex were
taken into account through the Doppler shift of quasiparticle
Hamiltonian, and it was shown that they do not lead to qualitative
changes, but give rise to important quantitative corrections to
the vortex mass. Vortex dynamics was studied both perturbatively and non-perturbatively, and the two approaches yielded the same results. The perturbation theory was formulated in a dual description of d-wave superconductors, where the Cooper-pair fluctuations are handled as a U(1) gauge field that mediates interactions between ``charged'' particle-like vortices. The non-perturbative approach was focused on the small-amplitude response of a vortex to a weak oscillating force. Our findings are also in agreement with simple scaling arguments that regard the d-wave quasiparticles as a quantum critical system.

Our main conclusions are the following: (1) at zero temperature
the vortex mass is finite and completely specifies vortex
dynamics, apart from a universal ``sub-Ohmic'' damping; (2) at finite temperatures vortex motion becomes
dissipated in an Ohmic fashion, with certain logarithmic
corrections. Furthermore, quasiparticles mediate new interactions between
vortices (assumed to be far enough apart so that quasiparticle
motion between vortices is incoherent), leading to both
longitudinal and transversal forces in addition to the usual
``hydrodynamic'' forces. All parameters that characterize vortex
dynamics depend quantitatively (and not qualitatively) on the
boundary conditions at the vortex core. We reached these
conclusions under several assumptions. First, the vortex core was
idealized by a point in the continuum limit. This was primarily
motivated by the fact that in good $d$-wave superconductors the
coherence length $\xi$ is not much larger than the Fermi
wavelength $k_F^{-1}$, so that the core is indeed small.
Furthermore, the superconducting gap vanishes in the nodal
momentum directions, leaving a route for quasiparticles inside the
core to leak out. Both of these facts conspire against bound core
states, which is the situation that our idealization of the core
models well, and various other studies support. The vortex mass
is, then, determined by scattering of free quasiparticles from a
moving vortex. Smallness of the core renders quantum phenomena
very important. At finite temperatures we ignored thermal
fluctuations of the superfluid order parameter, but kept the
quasiparticles in thermal equilibrium. Presumably, as long as
temperatures are low and quasiparticle contributions to vortex
dynamics dominant, such an  approximation is valid. Effects of
thermal non-equilibrium have not been taken into account, but they
are expected not to be important in the circumstances that we
consider (slow oscillatory vortex motion with very small
amplitude).

As we noted in Section~\ref{sec:intro}, there is a possible
disagreement between our results and the semiclassical
approach\cite{volovik,KopVin,KopVin2,kopnin}. The latter works find
that the nodal quasiparticles induce an effective mass which
diverges linearly with the separation between vortices. In
particular, it does appear that direct application of our theory and
the semiclassical theory leads to distinct predictions for a certain
class of experiments. It has recently been
argued\cite{CompOrd1,EstVMass} that the zero point quantum motion of
vortices leads to modulations in the local density of states, and
this was proposed as the explanation of recent STM experiments on
the cuprate superconductors \cite{hoffman,fischer}. The spatial
extent of the zero point motion, and hence density modulations, is
determined \cite{EstVMass} by the value of $m_{\Tr{v}}$. (Indeed,
this fact was the initial motivation for our study.) Consequently,
$m_{\Tr{v}}$ can be determined, in principle, from STM measurements
of the modulations. Our theory predicts that such an $m_{\Tr{v}}$
should be independent of the field $H$, at small $H$, while the
semiclassical theory predicts a divergence in $m_{\Tr{v}}$ as $\sim
1/\sqrt{H}$.

In the future, it is also possible that dynamic (THz scale)
measurements of vortex dynamics will lead to more direct
quantitative determinations of $m_{\Tr{v}}$, and these should allow
tests of the theoretical proposals.

At the end, we make some comments about the incoherent quasiparticle
mediated interactions between vortices, explored in the Appendix \ref{ssecVInt}.
They were modeled by an
effective chemical potential $\mu$ in the linearized quasiparticle
Hamiltonian, which describes the Doppler shift due to the
supercurrents from a second nearby vortex. The interactions that we
discussed in Appendix~\ref{ssecVInt} are specific to $d$-wave
superconductors (for similar effects to be observed in $s$-wave
superconductors, $\mu$ would have to be larger than the gap, that is
the two vortices practically overlapping). The effects of these
interactions would be visible in vortex-vortex scattering events:
the vortices would lose some of their initial kinetic energy during
scattering, and excite some low-energy extended quasiparticle
states. Anisotropy of $d$-wave superconductors would be reflected in
these interactions, so that the differential scattering
cross-section, and the amount of kinetic energy lost, would depend
on the initial direction of approach toward the point of collision.
We also speculate that these interactions might play an important
role in shaping the properties of vortex lattices (especially if
pinning disorder ruins the vortex lattice inversion symmetry). They
could lead to anisotropic stiffness and pinning of the vortex
lattice orientation to that of the substrate. The discovered
dissipation-looking response of a single vortex in presence of
another vortex could transform into significant and magnetic field
dependent correction to the vortex mass in the equilibrium
conditions of the vortex lattice. We leave these interesting issues
for future work.

\section{Acknowledgements}

We are very grateful to Zlatko Te\v sanovi\'c for several
stimulating discussions, his insightful perspective on the
semiclassical approach, and for freely sharing the results of
Ref.~\onlinecite{Tesh1} prior to publication. We also thank Matthew
Fisher for explaining aspects of Ref.~\onlinecite{bf} to us. We
acknowledge useful discussions with E.~Demler, B.~I.~Halperin, and
A.~Kolezhuk. This research was supported by NSF Grant DMR-0537077.

\appendix

\section{Details of the vortex action derivation}\label{appVortexAction}

In this appendix we provide support for the section
\ref{secVortexAction} by explicitly evaluating traces from the
expansion ~(\ref{VAct2}). At the first order in vortex displacement
we have:
\begin{eqnarray}
& - & \Tr{tr} \left\lbrack G_0 (\Bf{r}_{\Tr{v}} \nabla_{\Tr{v}}) H_0 \right\rbrack = \nonumber \\
& = & -\int \dd\tau\dd^2 r \ \Tr{tr} \left\lbrack G_0(\Bf{r},\tau;\Bf{r},\tau)
  (\Bf{r}_{\Tr{v}}(\tau) \nabla_{\Tr{v}}) H_0(\Bf{r}) \right\rbrack \nonumber \\
& = & \int \dd\tau\dd^2 r \sum_{n} \frac{1}{\beta} \sum_{\omega}
  \frac{e^{i\omega 0^+}}{-i\omega + \epsilon_n} \times \nonumber \\
& & \Tr{tr} \left\lbrack
  \psi_n^{\phantom{\dagger}}(\Bf{r}) \psi_n^{\dagger}(\Bf{r})
  \Bf{r}_{\Tr{v}}(\tau) \nabla H_0(\Bf{r}) \right\rbrack \\
& = & \int \dd\tau\dd^2 r \sum_{n} \frac{1}{\beta} \sum_{\omega}
  \frac{e^{i\omega 0^+}}{-i\omega + \epsilon_n} (\epsilon_n - \epsilon_n) \times \nonumber \\
& &  \Tr{tr} \left\lbrack
  \left( \Bf{r}_{\Tr{v}}(\tau) \nabla \psi_n^{\phantom{\dagger}}(\Bf{r}) \right)
  \psi_n^{\dagger}(\Bf{r}) \right\rbrack = 0 \ , \nonumber
\end{eqnarray}
where we have used  ~(\ref{Green0}), ~(\ref{Schr}), and integration by parts. For integration purposes, in absence of disorder, we may replace the vortex position gradient $\nabla_{\Tr{v}}$  with the negative normal gradient: $\nabla_{\Tr{v}} \to -\nabla$. This is just a shift of the coordinate system that puts a displaced vortex back to the origin, and simplifies derivations.

At the second order in vortex displacement we have two contributions. One comes from the second term in ~(\ref{VAct2}):
\begin{eqnarray}\label{Trap1}
& & -\frac{1}{2} \Tr{tr} \left\lbrack G_0 (\Bf{r}_{\Tr{v}} \nabla_{\Tr{v}})^2 H_0 \right\rbrack = \\
& & \quad -\frac{1}{2} \int \dd\tau \dd^2 r \dd^2 r' \sum_{n,n'} \frac{1}{\beta} \sum_{\omega}
  \frac{e^{i\omega 0^+}}{-i\omega + \epsilon_n} \times \nonumber \\
& & \quad \Tr{tr} \left\lbrack \psi_n^{\phantom{\dagger}}(\Bf{r}) \psi_n^{\dagger}(\Bf{r}')
  (\Bf{r}_{\Tr{v}}(\tau) \nabla)^2
  \left( \epsilon_{n'} \psi_{n'}^{\phantom{\dagger}}(\Bf{r}')
  \psi_{n'}^{\dagger}(\Bf{r}) \right) \right\rbrack \ . \nonumber
\end{eqnarray}
In order to evaluate it, we sum over the frequencies:
\begin{equation}\label{OmegaSum1}
\frac{1}{\beta} \sum_{\omega} \frac{e^{-i\omega\tau}}{-i\omega+\epsilon} =
  e^{-\epsilon\tau} \left\lbrace
  \begin{array}{lcl}
    -f(\epsilon)  & \quad , \quad & \tau<0 \\
    1-f(\epsilon) & \quad , \quad & \tau>0
  \end{array}
\right\rbrace \ ,
\end{equation}
Then, after integration by parts, some algebraic manipulation and making use of the fact that the wavefunctions $\psi_n(\Bf{r})$ form a complete set of states, the expression ~(\ref{Trap1}) reduces to:
\begin{equation}\label{Trap2}
\frac{1}{2} \sum_{n,n'} \left( f(\epsilon_n) - f(\epsilon_{n'}) \right)
  (\epsilon_{n'} - \epsilon_{n}) \int\dd\tau | \Bf{r}_{\Tr{v}}(\tau) \Bf{U}_{n,n'} |^2
\end{equation}
This has the form of a trapping potential.

The other contribution to the second order in vortex displacement comes from the third term in ~(\ref{VAct2}):
\begin{eqnarray}\label{appTrace3a}
& & \frac{1}{2} \Tr{tr} (G_0 \Delta H)^2 = \\
& &  \int \dd\tau \dd^2 r \int \dd\tau' \dd^2 r'
  \Tr{tr} \Bigl\lbrack G_0(\Bf{r},\tau;\Bf{r}',\tau')
  \left( \Bf{r}_{\Tr{v}}(\tau') \nabla_{\Tr{v}} H_0(\Bf{r}') \right) \nonumber \\
& & \qquad  G_0(\Bf{r'},\tau';\Bf{r},\tau)
  \left( \Bf{r}_{\Tr{v}}(\tau) \nabla_{\Tr{v}} H_0(\Bf{r}) \right) \Bigr\rbrack \nonumber \\
&  & = -\frac{1}{2} \int \dd\tau \dd\tau' \sum_{n,n'} \frac{1}{\beta^2} \sum_{\omega,\omega'}
  \frac{e^{-i(\omega-\omega')(\tau-\tau')}}{(-i\omega+\epsilon_n)(-i\omega'+\epsilon_{n'})} \times \nonumber \\
& & \qquad  (\epsilon_n - \epsilon_{n'})^2
  \left( \Bf{r}_{\Tr{v}}(\tau') \Bf{U}_{n,n'} \right)
  \left( \Bf{r}_{\Tr{v}}(\tau) \Bf{U}_{n',n} \right) \ . \nonumber
\end{eqnarray}
The matrix elements $\Bf{U}_{n,n'}$ are defined by ~(\ref{UMatrEl}). If we integrate out $\tau$ and $\tau'$ and use $\Bf{U}_{n',n} = -\Bf{U}_{n,n'}^*$ we obtain:
\begin{eqnarray}\label{appTrace3b}
\frac{1}{2} \Tr{tr} (G_0 \Delta H)^2 & = &
  \frac{1}{2} \sum_{n,n'} \frac{1}{\beta^2} \sum_{\omega,\omega'}
  \frac{e^{-i(\omega-\omega')(\tau-\tau')}}{(-i\omega+\epsilon_n)(-i\omega'+\epsilon_{n'})} \times \nonumber \\
& &  (\epsilon_n - \epsilon_{n'})^2
  | \Bf{r}_{\Tr{v}}(\omega-\omega') \Bf{U}_{n,n'} |^2 \ .
\end{eqnarray}
Let $\Delta\omega = \omega - \omega'$. We can carry out summation over $\omega$ using the formula:
\begin{equation}
\frac{1}{\beta} \sum_{\omega} \frac{1}{(-i\omega+\epsilon_1)(-i\omega+\epsilon_2)} =
  \frac{f(\epsilon_1)-f(\epsilon_2)}{\epsilon_1-\epsilon_2} \ .
\end{equation}
The trace ~(\ref{appTrace3b}) becomes:
\begin{eqnarray}\label{appTrace3c}
\frac{1}{2} \Tr{tr} (G_0 \Delta H)^2 & = &
  \frac{1}{2} \sum_{n,n'} \frac{1}{\beta} \sum_{\Delta\omega}
  \frac{f(\epsilon_n)-f(\epsilon_{n'}+i\Delta\omega)}
  {\epsilon_n-\epsilon_{n'}-i\Delta\omega} \times \nonumber \\
& & (\epsilon_n - \epsilon_{n'})^2
  | \Bf{r}_{\Tr{v}}(\Delta\omega) \Bf{U}_{n,n'} |^2 \ .
\end{eqnarray}
Note that $\Delta\omega$ takes values $2\pi T \times \textrm{integer}$, so that $f(\epsilon_{n'}+i\Delta\omega)$ is simply $f(\epsilon_{n'})$.

The vortex action at the second order in vortex displacement is obtained by adding ~(\ref{Trap2}) and ~(\ref{appTrace3c}). In expression ~(\ref{Trap2}) we perform Fourier transformation of $\Bf{r}_{\Tr{v}}(\tau)$ and integrate over $\tau$, while in expression ~(\ref{appTrace3c}) we relabel $\Delta\omega$ into $\omega$. The result is given by ~(\ref{VActFull}).

\section{Notes on the derivation of $\Bf{U}_{n_1,n_2}$ in absence of the Doppler shift}\label{appUMatrEl}

Here we outline some steps that lead to the result ~(\ref{UExpr}).
We calculate ~(\ref{U2}) in the rescaled coordinate system upon
substitution of ~(\ref{WF}) and ~(\ref{WF0}), together with the
representation of gradient in the polar coordinates
~(\ref{DerPol}). In order to treat all cases of angular momentum
on the same footing in ~(\ref{WF}) and ~(\ref{WF0}), we label the
indices of the Bessel functions in the upper and lower spinor
component by $l'$ and $l''$ respectively. Later we worry about
dependence of $l'$ and $l''$ on $l$. First we integrate out the
polar angle $\phi$. This prohibits all transitions between states
whose angular momenta do not differ by one. Hence, let $\sigma =
l_2 - l_1 = \pm 1$. Then, we use the following identities for the
Bessel functions:
\begin{eqnarray}\label{BesselIdentities}
\frac{l}{kr} J_l(kr) & = & \frac{1}{2} \left( J_{l-1}(kr) + J_{l+1}(kr) \right) \\
\frac{1}{k}\frac{\partial}{\partial_r} J_l(kr)
  & = & \frac{1}{2} \left( J_{l-1}(kr) - J_{l+1}(kr) \right) \nonumber
\end{eqnarray}
in order to carry out differentiation over $r$ and eliminate the $1/r$ factors that originate from the azimuthal gradient components. We introduce notation:
\begin{equation}\label{Gamma}
\gamma_{l_1,l_2}(k_1,k_2) = \int\limits_0^{\infty} \dd r r J_{l_1}(k_1 r) J_{l_2}(k_2 r) \ ,
\end{equation}
and
\begin{eqnarray}\label{AA}
A' & = & \frac{1}{2} \sqrt{\frac{k_1 k_2}{|E_1 E_2|}}
  \left( \sqrt{E_1 - m} \right)^* \sqrt{E_2 - m} \\
A'' & = & q_1 q_2
  \Tr{sign} \left( l_1 - \frac{1}{2} \right) \Tr{sign} \left( l_2 - \frac{1}{2} \right) \times \nonumber \\
& &  \frac{1}{2} \sqrt{\frac{k_1 k_2}{|E_1 E_2|}}
  \left( \sqrt{E_1 + m} \right)^* \sqrt{E_2 + m} \nonumber \ .
\end{eqnarray}
After some algebraic manipulations, we arrive at:
\begin{eqnarray}\label{U3}
\Bf{U}_{n_1,n_2} & = &
  \left( \frac{\hat{\Bf{x}}}{v_F} + i\sigma \frac{\hat{\Bf{y}}}{v_{\Delta}} \right)
  \frac{k_2}{4} \times \\
& & \Big\lbrack
       A' \left( 1 + \sigma \frac{l_2-1}{l'_2} \right) \gamma_{l'_1,l'_2-1}(k_1,k_2) + \nonumber \\
& & A' \left( -1 + \sigma \frac{l_2-1}{l'_2} \right) \gamma_{l'_1,l'_2+1}(k_1,k_2) + \nonumber \\
& & A'' \left( 1 + \sigma \frac{l_2}{l''_2} \right) \gamma_{l''_1,l''_2-1}(k_1,k_2) + \nonumber \\
& & A'' \left( -1 + \sigma \frac{l_2}{l''_2} \right) \gamma_{l''_1,l''_2+1}(k_1,k_2)
       \Big\rbrack \ , \nonumber
\end{eqnarray}
for $\sigma = l_2 - l_1 = \pm 1$ (otherwise, $\Bf{U}_{n_1,n_2}=0$). Next, we evaluate $\gamma_{l_1,l_2}(k_1,k_2)$. Note that for the case of interest $\theta=0$ we can write:
\begin{eqnarray}
l' & = & \left\lbrace
  \begin{array}{lcl}
    -l + \frac{1}{2} & \quad , \quad & l \leq 0 \\
    l - \frac{1}{2} & \quad , \quad & l > 0
  \end{array}
\right\rbrace \\[2mm]
l'' & = & \left\lbrace
  \begin{array}{lcl}
    -l - \frac{1}{2} & \quad , \quad & l \leq 0 \\
    l + \frac{1}{2} & \quad , \quad & l > 0
  \end{array}
\right\rbrace \ . \nonumber
\end{eqnarray}
If this is substituted in ~(\ref{U3}), than it can be easily seen that the angular momentum indices of $\gamma$ never differ by more than two. This allows us to evaluate the $\gamma$ terms. If the angular momentum indices are equal, then:
\begin{equation}\label{GammaNorm}
\gamma_{l,l}(k_1,k_2) = \frac{1}{k_2} \delta(k_2 - k_1) \ .
\end{equation}
If the angular momentum indices differ by two, then use the first identity from ~(\ref{BesselIdentities}) to obtain:
\begin{eqnarray}
\gamma_{l,l \pm 2}(k_1,k_2) & = & \frac{2(l \pm 1)}{k_2} \int\limits_0^{\infty}\dd r
  J_l(k_1 r) J_{l \pm 1}(k_2 r) \nonumber \\
  & - & \gamma_{l,l}(k_1,k_2) \ .
\end{eqnarray}
The integral above is a special case of Weber-Schafheitlin formula
\cite{AbrStegun}, which yields simple expressions:
\begin{eqnarray}
\gamma_{l,l-2}(k_1,k_2) & = & \frac{2(l-1)}{k_2^2} \left( \frac{k_2}{k_1} \right)^l
  \Theta(k_1-k_2) \nonumber \\
  & - & \frac{1}{k_2} \delta(k_1-k_2) \\
\gamma_{l,l+2}(k_1,k_2) & = & \frac{2(l+1)}{k_1^2} \left( \frac{k_1}{k_2} \right)^{l+2}
  \Theta(k_2-k_1) \nonumber \\
  & - & \frac{1}{k_1} \delta(k_2-k_1) \ , \nonumber
\end{eqnarray}
where $\Theta(x)$ is a step-function:
\begin{equation}
\Theta(x) = \left\lbrace
  \begin{array}{lcl}
    1 & \quad , \quad & x>0 \\
    \frac{1}{2} & \quad , \quad & x=0 \\
    0 & \quad , \quad & x<0
  \end{array}
\right\rbrace \ .
\end{equation}
If $l_1=0$ or $l_2=0$ then in some cases the two angular momentum indices of the $\gamma$ terms may differ by one. Then, however, exact expressions for the appropriate Bessel functions are simple enough to directly integrate out $\gamma$ and add their contributions to $\Bf{U}_{n_1,n_2}$. Further manipulations are still quite complex, and we do not include them here. We only note that it is safe at this stage to set $m=0$ (no singularities are introduced). The final result is ~(\ref{UExpr}).

\section{Calculation of the vortex mass at zero temperature in absence of the Doppler shift}\label{appVortexMass}

In this appendix we substitute ~(\ref{USumL}) into ~(\ref{VortexMass}) term by term, and explicitly calculate the integral for the vortex mass in absence of the Doppler shift. The goal is to show that there are no infra-red divergences, and to provide an estimate of the vortex mass in terms of the electron mass.

The first term of ~(\ref{USumL}) contributes the following to the vortex mass:
\begin{eqnarray}
M^{(1)}_{\Tr{v}} & \propto &
  \int\limits_{-\Lambda}^{0}\dd\epsilon_1 \int\limits_{0}^{\Lambda}\dd\epsilon_2
    \frac{1}{\epsilon_2-\epsilon_1}
  \frac{(\epsilon_1+\epsilon_2)^2 (\epsilon_1^2+\epsilon_2^2)}
     {|\epsilon_1^2-\epsilon_2^2| \Tr{max}(\epsilon_1^2,\epsilon_2^2)} \nonumber \\
& = &  \int\limits_{0}^{\Lambda} \dd k_1 \int\limits_{0}^{\Lambda} \dd k_2 \frac{1}{k_2 + k_1}
  \frac{(k_2-k_1)^2 (k_1^2+k_2^2)}
     {|k_2^2-k_1^2| \Tr{max}(k_1^2,k_2^2)} \nonumber \\
& = & 2 \int\limits_{0}^{\Lambda}\dd k_1 \int\limits_{0}^{k_1}\dd k_2
  \frac{(k_1-k_2)(k_1^2+k_2^2)}{(k_1+k_2)^2 k_1^2} = \cdots \nonumber \\
&= & 3(3-4\log2)\Lambda = 0.682234 \Lambda \ , \nonumber
\end{eqnarray}
where the constant of proportionality is:
\begin{equation}\label{MassConst}
C = \frac{1}{16} \left( \frac{1}{v_F^2} + \frac{1}{v_{\Delta}^2} \right) \ .
\end{equation}
This contribution to the vortex mass includes effects of all particle-hole transitions between states that do not involve the zero angular momentum channel. No infra-red divergency has been encountered. The second term of ~(\ref{USumL}) includes partially the effects of transitions that involve the zero angular momentum channel. This term is the most divergent as $E_2-E_1 \to 0$, and thus the most dangerous. However, it only gives a finite contribution to the vortex mass:
\begin{eqnarray}
M^{(2)}_{\Tr{v}} & \propto &
  \int\limits_{-\Lambda}^{0}\dd\epsilon_1 \int\limits_{0}^{\Lambda}\dd\epsilon_2
    \frac{1}{\epsilon_2-\epsilon_1} \frac{8}{\pi^2}
    \left( \frac{\epsilon_1+\epsilon_2}{\epsilon_1-\epsilon_2} \right)^2 \nonumber \\
& = & \frac{8}{\pi^2} \int\limits_{0}^{\Lambda} \dd k_1 \int\limits_{0}^{\Lambda} \dd k_2
    \frac{(k_2-k_1)^2}{(k_2+k_1)^3} \nonumber \\
& = & \frac{2}{\pi^2} \int\limits_{0}^{\Lambda}\dd\Delta k
       \int\limits_{\Delta k / 2}^{\Lambda - \Delta k / 2}\dd k \frac{\Delta k^2}{k^3}
  = \cdots \nonumber \\
& = & \frac{16}{\pi^2} \left( \log2-\frac{1}{2} \right)\Lambda
  = 0.313118 \Lambda \ . \nonumber
\end{eqnarray}
The rest of ~(\ref{USumL}) is hard to integrate exactly, but it is easy to show that there are no infra-red divergences:
\begin{eqnarray}
M^{(3)}_{\Tr{v}} & \propto & \frac{\Lambda}{\pi^2}
  \int\limits_{0}^{1}\dd x_1 \int\limits_{0}^{1}\dd x_2
  \frac{(x_1-x_2)^2}{x_1+x_2} \times \nonumber \\
& & \hspace{-15.5mm} \left\lbrack
    \frac{1}{4} \left( \frac{1}{x_1^2} + \frac{1}{x_2^2} \right)
    \log^2 \left( \frac{x_1-x_2}{x_1+x_2} \right)^2
    + \frac{2}{x_1 x_2} \log \left( \frac{x_1-x_2}{x_1+x_2} \right)^2 \right\rbrack \nonumber \\
& = & \frac{2\Lambda}{\pi^2} \int\limits_0^{\pi/4}\dd \phi \int\limits_0^{1/\cos\phi} r \dd r
  \frac{1}{r} \frac{1-\sin 2\phi}{\cos\phi+\sin\phi} \times \nonumber \\
& & \hspace{-15.5mm} \left\lbrack
  \frac{4}{\sin 2\phi} \log \left( \frac{1-\sin 2\phi}{1+\sin 2\phi} \right)
  + \frac{1}{\sin^2 2\phi} \log^2 \left( \frac{1-\sin 2\phi}{1+\sin 2\phi} \right)
  \right\rbrack \nonumber \\
& = & \frac{2\Lambda}{\pi^2} \int\limits_0^{\pi/4}\dd \phi
  \frac{1}{\cos\phi} \frac{1-\sin 2\phi}{\cos\phi+\sin\phi} \times \nonumber \\
& & \hspace{-15.5mm} \left\lbrack
  \frac{4}{\sin 2\phi} \log \left( \frac{1-\sin 2\phi}{1+\sin 2\phi} \right)
  + \frac{1}{\sin^2 2\phi} \log^2 \left( \frac{1-\sin 2\phi}{1+\sin 2\phi} \right)
  \right\rbrack \nonumber \\[2mm]
& = & \cdots = -0.196462 \Lambda \ , \nonumber
\end{eqnarray}
where $x_1 = k_1/\Lambda = r\cos\phi$, $x_2 = k_2/\Lambda = r\sin\phi$, and symmetry with respect to $x_1 \leftrightarrow x_2$ was used. It is easy to check that the function of $\phi$ inside the last integral is finite for every $\phi \in (0,\pi/4)$. The total vortex mass is:
\begin{equation}
m_{\Tr{v}} = M^{(1)}_{\Tr{v}} + M^{(2)}_{\Tr{v}} + M^{(3)}_{\Tr{v}}
  \approx 0.05 \left( \frac{1}{v_F^2} + \frac{1}{v_{\Delta}^2} \right) \Lambda \ .
\end{equation}

\section{Quasiparticle mediated interactions between
vortices}\label{ssecVInt}

In this appendix we will examine how the gapless quasiparticles of a
$d$-wave superconductor mediate certain interactions between
vortices. Consider placing a second vortex near the vortex whose
dynamics we want to study. We assume here that the second vortex is
far enough so that the fermionic quasiparticles near the two
vortices can be treated independently. This requires some external
source of dissipation, which will render quasiparticle motion
between the vortices incoherent. We assume in the remainder of this
appendix that such a source of dissipation is present, and so our
conclusions here are limited to this case.

Recall the full linearized Hamiltonian for quasiparticles
~(\ref{BdG}). Presence of a second vortex affects the quasiparticles
that move around the first vortex in two ways. First, the ``gauge''
field $\Bf{a}(\Bf{r})$ is changed, but this is not crucial because
we can always ``gauge'' it away. Second, the superfluid velocity
$\Bf{v}(\Bf{r})$ is changed, which modifies the Doppler shift term.
If the distance between two vortices is large, the supercurrent due
to the second vortex appears practically uniform in vicinity of the
first vortex. The corresponding Doppler shift formally looks like a
chemical potential $\mu$ presented to the linearized quasiparticles.
This effective chemical potential depends on the distance between
vortices, and their orientation with respect to the nodal directions
($\mu \propto v_x$, where $v_x$ is the component of superfluid
velocity in the nodal direction $x$).

\begin{figure}
\includegraphics[width=2.5in]{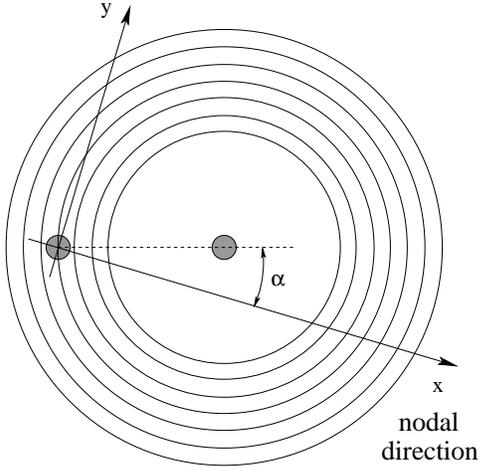}
\caption{\label{vortex-interaction}One vortex in vicinity of
another. Supercurrents due to the second vortex appear roughly
uniform in the region occupied by the first vortex, resulting with a
uniform Doppler shift of quasiparticle energies.}
\end{figure}

In order to explore qualitative aspects of these interactions, we
ignore complications arising from the vortex self-Doppler shift and
non-uniformity of the supercurrents. We simply add a chemical
potential $\mu I$ to the Hamiltonian ~(\ref{HNoDoppler}). Effect of
this is that all energies are shifted by $\mu$ in the vortex action.
We restrict our attention only to zero temperature, and assume $\mu
> 0$ without loss of generality. Consider first the ``longitudinal
correlations'':
\begin{eqnarray}
F_{\parallel}(\omega) & \propto &
  \int\limits_{-\Lambda}^{\mu}\dd E_1 \int\limits_{\mu}^{\Lambda}\dd E_2
  \frac{\omega^2 (E_2 - E_1)}{(E_2 - E_1)^2 + \omega^2} \times \nonumber  \\
& &  \sum_{\Tr{nodes}}\sum_{l,\sigma} | \Bf{r}_{\Tr{v}}(\omega)
\Bf{U}_{n_1,n_2} |^2 \ .
\end{eqnarray}
Similar to the case of finite temperatures and $\mu=0$, certain
``dangerous'' particle-particle transitions are allowed when
$\mu>0$. They occur at $E_1 = E_2 = \mu$, where the bulk LDOS is not
zero. Following the procedure from subsection \ref{ssecFiniteT}, we
find that the transitions involving the zero angular momentum
channel dominate at small frequencies:
\begin{equation}
F_{\parallel}(\omega) \propto \mu^2 |\omega| \quad , \quad |\omega|
\ll \mu \ .
\end{equation}
The first correction at small frequencies comes from the transitions
that do not involve the zero angular momentum channel:
\begin{equation}
\Delta F_{\parallel}(\omega) \propto \mu \omega^2 \log \left(
\frac{\mu}{|\omega|} \right) \quad , \quad |\omega| \ll \mu \ .
\end{equation}
Clearly, the anomalous dynamics at small frequencies and non-zero
effective chemical potential is of the same kind as that found in
the case of finite temperature. Only the energy scale that controls
it has been changed.

The remaining corrections are $\mathcal{O}(\omega^2)$, meaning that
vortex inertia is also affected by the presence of another vortex.
Physical consequences of all these effects could be directly
observed in vortex-vortex scattering events as vortex kinetic energy
loss to the quasiparticles. Also, dynamics of a vortex lattice
should be affected. Due to quantum fluctuations of vortices the quasiparticle mediated forces
would tend to pin the orientation of the vortex lattice to the
substrate, and they would influence its stiffness. These effects
would be even greater in absence of perfect inversion symmetry of
the vortex lattice, caused for example by vortex pinning disorder.

Now consider the ``transversal correlations'':
\begin{eqnarray}
F_{\perp}(\omega) & \propto &
  \int\limits_{-\Lambda}^{\mu}\dd E_1 \int\limits_{\mu}^{\Lambda}\dd E_2
  \frac{-i\omega (E_2 - E_1)}{E_2 - E_1 - i\omega} \times \nonumber \\
& &   \sum_{l,\sigma} \sigma | U_{(q_1,l-\sigma,k_1),(q_2,l,k_2)}
|^2 \ .
\end{eqnarray}
As shown in subsection \ref{ssecTransDyn}, the symmetry under $E_1
\to -E_2$ and $E_2 \to -E_1$ would set $ F_{\perp}(\omega)=0$.
However, this symmetry is removed by finite $\mu$. By substituting
~(\ref{USumT}) we find:
\begin{equation}
F_{\perp}(\omega) \propto -i\omega \left( \mu\Lambda +
\mathcal{O}(\mu^2\Tr{sign}(\mu)) \right)
  + \mathcal{O}(\omega^3) \ .
\end{equation}
This result can be qualitatively understood by noting that the
formal particle-hole symmetry of the linearized quasiparticle
Hamiltonian is broken when $\mu \neq 0$. There is a net
quasiparticle current density, formed by occupied states within the
energy range $0<E<\mu$. This is possible because the exact
wavefunctions ~(\ref{WF}) and ~(\ref{WF0}) are mixtures of a
particle and hole that move with different angular velocities.

\newpage

\section{Other values of the core parameter $\theta$}\label{appTheta}

All calculations in Section~\ref{secAnalytic} were restricted to
boundary condition at the vortex core $\theta=0$, which physically
corresponds to a complete enclosure of the vortex flux within a
finite-sized cylinder. Here we briefly discuss other boundary
conditions. For general value of $\theta$ the transition matrix
elements ~(\ref{UExpr}) are modified at $l=(\sigma \pm 1)/2$:
\begin{widetext}
\begin{eqnarray}
U_{(q_1,l-\sigma,k_1),(q_2,l,k_2)} & = & \cos(\theta) \left\lbrack
      \frac{2 \sigma q_1 q_2}{\pi}\frac{E_1+E_2}{E_1-E_2} +
      \frac{C_{\sigma}}{2\pi}
      \left( \frac{k_1}{k_2} \right)^{\sigma l - \frac{1}{2}}
      \frac{E_1 + E_2}{\sqrt{k_1 k_2}}
      \log \left( \frac{k_1-k_2}{k_1+k_2} \right)^2 \right\rbrack + \nonumber \\
& &  \sin(\theta) \left\lbrack
      4 \sqrt{k_1 k_2} \delta(E_2 - E_1) -
      C_{\sigma} \left( \frac{k_1}{k_2} \right)^{\sigma l - \frac{1}{2}}
      \frac{E_1 + E_2}{\sqrt{k_1 k_2}}
      \Theta \left( \sigma (k_2 - k_1) \right) \right\rbrack \quad , \quad l = \frac{\sigma+1}{2} \nonumber \\
U_{(q_1,l-\sigma,k_1),(q_2,l,k_2)} & = & -\sin(\theta) \left\lbrack
      \frac{2 \sigma q_1 q_2}{\pi} \left( q_1 q_2 \frac{E_1+E_2}{E_1-E_2} + 2 q_1 q_2 \sigma \right) +
      \frac{C_{\sigma}}{2\pi}
      \left( \frac{k_1}{k_2} \right)^{\sigma l - \frac{1}{2}}
      \frac{E_1 + E_2}{\sqrt{k_1 k_2}}
      \log \left( \frac{k_1-k_2}{k_1+k_2} \right)^2 \right\rbrack + \nonumber \\
& &  \cos(\theta) \left\lbrack
      - 4 \sqrt{k_1 k_2} \delta(E_1 - E_2) -
      C_{\sigma} \left( \frac{k_1}{k_2} \right)^{\sigma l - \frac{1}{2}}
      \frac{E_1 + E_2}{\sqrt{k_1 k_2}}
      \Theta \left( \sigma (k_1 - k_2) \right) \right\rbrack \quad , \quad l = \frac{\sigma-1}{2} \nonumber
\end{eqnarray}
\end{widetext}
No new singularities have been introduced, and all results from
previous subsections remain qualitatively the same. However, there
are quantitative changes to the vortex mass, finite temperature
dissipation coefficient, etc. Therefore, the parameters that
characterize vortex dynamics depend on the boundary conditions at
the vortex core. We will not investigate further this dependence.

\section{Representation of the full quasiparticle Hamiltonian}\label{appDoppler}

Here we construct the representation of the Hamiltonian ~(\ref{BdG}) in the basis of states ~(\ref{WF}), ~(\ref{WF0}). This is used to numerically diagonalize the Hamiltonian on a finite sample with radius $R$.  The basis wavefunctions ~(\ref{WF}) and ~(\ref{WF0}) are easily renormalized to unity on the finite sample if multiplied by $\sqrt{\pi / R}$. With an appropriate hard-wall boundary condition momentum is quantized to zeros of the Bessel's functions:
\begin{equation}\label{DescrK}
k = | E | = \frac{\pi}{2R}
  \left( \left\vert l-\frac{1}{2} \right\vert + 2n - \frac{1}{2} \right)
  \quad , \quad kR \gg 1 \ .
\end{equation}
Note that the degeneracy with respect to angular momentum $l$ is lifted in a finite system. In the chosen representation, the full Hamiltonian ~(\ref{BdG}) is a sum of a diagonal matrix that contains discretized energies $E$ and the Doppler shift part whose matrix elements $\la E_1, l_1 | H_D | E_2, l_2 \ra$ can be evaluated using the rescaled coordinate system introduced in section \ref{ssecMatrEl}. Assuming that the supercurrent $\Bf{v}$ describes a $\pi$-flux localized at the origin, as discussed in the section \ref{secHamiltonian}, we have (at the $\Bf{p} = \hbar k_F \hat{\Bf{x}}$ node):
\begin{eqnarray}
& & \hspace{-6mm} \la E_1, l_1 | H_D | E_2, l_2 \ra = -\frac{v_F v_{\Delta}}{2}
  \int\limits_0^{2\pi}\dd\phi \int\limits_0^R\dd r \\
& &  \frac{\sin\phi}{(v_F \cos\phi)^2 + (v_{\Delta} \sin\phi)^2}
  \psi^{\dagger}_{E_1,l_1}(r,\phi) \psi^{\phantom{\dagger}}_{E_2,l_2}(r,\phi) \nonumber \\
& = & -\frac{v_F v_{\Delta}}{2} \frac{(\sqrt{E_1})^*\sqrt{E_2}}{4R}
  \Biggl\lbrack \overline{\gamma}_{l'_1,l'_2}(k_1,k_2) + \nonumber \\
& &  q_1 q_2 \Tr{sign}\left( l_1-\frac{1}{2} \right) \Tr{sign}\left( l_2-\frac{1}{2} \right)
  \overline{\gamma}_{l''_1,l''_2}(k_1,k_2) \Biggr\rbrack \times \nonumber \\
& &  \int\limits_0^{2\pi}\dd\phi
  \frac{\sin\phi \ e^{i(l_2-l_1)\phi}}{(v_F \cos\phi)^2 + (v_{\Delta} \sin\phi)^2} \ . \nonumber
\end{eqnarray}
where, like before, $q_i = \pm 1$, $E_i = q_i k_i$, and $l'$ and $l''$ are indices of the Bessel's functions in the upper and lower spinor components respectively of the wavefunctions ~(\ref{WF}) and ~(\ref{WF0}) (which depend on $l$). Recall that $\Bf{v}$ is gauge invariant, so that we cannot avoid its ``elliptical'' shape in the rescaled coordinate system. The integral over $\phi$ can be calculated exactly; for $l=l_2-l_1>0$ it is equal to:
\begin{equation}
  \frac{\pi}{v_{\Delta}} \frac{1}{v_F + v_{\Delta}}
  \left\vert \frac{v_F - v_{\Delta}}{v_F + v_{\Delta}} \right\vert^{\frac{l-1}{2}}
  \left\lbrace
  \begin{array}{clc}
    i^l - (-i)^l & \ , \ & v_F \geq v_{\Delta} \\
    i\left( 1 - (-1)^l \right) & \ , \ & v_F \leq v_{\Delta}
  \end{array}
  \right\rbrace . \nonumber
\end{equation}
The new function $\overline{\gamma}$ is:
\begin{equation}\label{GammaB}
\overline{\gamma}_{l_1,l_2}(k_1,k_2) = \int\limits_0^R\dd r
  J_{l_1}(k_1 r) J_{l_2}(k_2 r) \ .
\end{equation}
Analytic expression for $\overline{\gamma}$ is available only in the
$R\to\infty$ limit (a special case of Weber-Schafheitlin formula):
\begin{widetext}
\begin{eqnarray}\label{GammaBar}
\overline{\gamma}_{l_1,l_2}(k_1,k_2) & \approx & \left\lbrace
  \begin{array}{lcl}
    \frac{k_2^{l_2}}{k_1^{l_2+1}} \frac{\Gamma\left(\frac{l_1+l_2+1}{2}\right)}
      {\Gamma(l_2+1)\Gamma\left(\frac{l_1-l_2+1}{2}\right)}
    \phantom{|}_2 F_1 \left( \frac{l_1+l_2+1}{2} , \frac{l_2-l_1+1}{2} ;
       l_2+1 ; \left(\frac{k_2}{k_1}\right)^2 \right) & \quad , \quad & k_1>k_2 \\
    \frac{k_1^{l_1}}{k_2^{l_1+1}} \frac{\Gamma\left(\frac{l_1+l_2+1}{2}\right)}
      {\Gamma(l_1+1)\Gamma\left(\frac{l_2-l_1+1}{2}\right)}
    \phantom{|}_2 F_1 \left( \frac{l_1+l_2+1}{2} , \frac{l_1-l_2+1}{2} ;
       l_1+1 ; \left(\frac{k_1}{k_2}\right)^2 \right) & \quad , \quad & k_2>k_1
  \end{array} \right\rbrace \ , \\[5mm]
\overline{\gamma}_{l_1,l_2}(k,k) & \approx & \left\lbrace
  \begin{array}{lcl}
    -\frac{1}{2k} \Tr{sign}\left(l_1+l_2+\frac{1}{2}\right)
      \Tr{sign}\left(l_1-l_2+\frac{1}{2}\right) (-1)^{\frac{l_1-l_2+1}{2}}
      & \quad , \quad & l_1-l_2 \Tr{ is an odd integer} \\
    \frac{1}{\pi k} \cos\left(\frac{l_2-l_1}{2}\pi\right) \log(kR)
      & \quad , \quad & \Tr{ otherwise in the limit } kR \gg 1
  \end{array} \right\rbrace \ . \nonumber
\end{eqnarray}
\end{widetext}

It turns out that the formulas ~(\ref{DescrK}) and ~(\ref{GammaBar}) provide a very good approximation for our purposes. This approximation scheme uses the fact that for all states of interest $kR \gg 1$ in the thermodynamic limit. The states with $kR \sim 1$ are not treated accurately when $R$ is finite, but all such states collapse to an infinitely small range of energies when $R \to \infty$. Then, the Hamiltonian representation can be constructed relatively fast. The results of diagonalization and vortex dynamics obtained within this approximation have been found virtually the same as the corresponding results with exact energy level discretization and exact numerical calculation of $\overline{\gamma}$ in finite systems.

\section{Semi-analytical calculation of $\mathcal{R}(\Delta\epsilon)$}\label{appSemiAnalytical}

Here we present a semi-analytical argument against dissipation and anomalous vortex dynamics caused by the Doppler shift. We make several assumptions about general properties of the exact wavefunctions $\Psi_{n}(E,l) = \la E,l | n \ra$ in the thermodynamic limit. First, we assume that the wavefunctions $\Psi_{n}(E,l)$ are analytic functions of $E$ everywhere except perhaps at $E = g(\epsilon)$, and also at $E=0$ and/or $\epsilon=0$. The function $g(\epsilon)$ satisfies:
\begin{eqnarray}\label{DScond}
g(\epsilon) & = & -g(-\epsilon) \\
\Tr{sign}(g(\epsilon)) & = & \Tr{sign}(\epsilon) \ , \nonumber
\end{eqnarray}
and gives the energy at which $\Psi_{n}(E,l)$ may be developing a Dirac peak in the thermodynamic limit. These properties of $g(\epsilon)$ are clearly visible from the numerical spectra (the first one is protected by a symmetry). The numerics essentially suggests $g(\epsilon)=\epsilon$, and this is in agreement with scattering theory in the thermodynamic limit. No divergences occur at $E=0$ or $\epsilon=0$, and these possible non-analyticities do not play an important role. The remaining assumptions follow from the requirement that the function $\mathcal{R}(\Delta\epsilon)$, which determines the vortex action, be finite for every finite $\Delta\epsilon$. All wavefunction singularities should be integrable; integrals of the wavefunction (not its modulus squared) across the point of possible non-analyticity are convergent in the thermodynamic limit:
\begin{equation}
\int\limits_{g(\epsilon)-\Delta E}^{g(\epsilon)+\Delta E}
  \hspace{-4mm} \dd E \hspace{1mm} | \Psi_{\epsilon,i}(E,l) | < \infty \ .
\end{equation}
This is certainly consistent with a possible Dirac peak. Also,
certain sums over angular momentum channels should be either finite
or lead to integrable singularities. All of these assumptions are
based on the numerically obtained wavefunctions and spectra.

We now analytically explore general properties of the function $\mathcal{R}(\Delta\epsilon)$ given by ~(\ref{RFunct}), and check whether any infra-red divergences might occur. A particular question is what the dependence on small $\Delta\epsilon$ is. If $\mathcal{R}(\Delta\epsilon) \sim \Delta\epsilon^{\nu}$ with $\nu \leq 0$, then vortex dynamics at small frequencies may be anomalous and not simply governed by the vortex inertia.

Let us introduce a convenient symbolic notation for the exact states $|n\ra = | \epsilon, i\ra$, where $\epsilon = \epsilon_n$ and $i$ is a fictitious quantum number that replaces the angular momentum $l$ of the states in absence of the Doppler shift. The purpose of this is only to explicitly reveal the role of energy $\epsilon$, which does not completely specify the exact states. Then we symbolically write ~(\ref{RFunct}) as:
\begin{eqnarray}\label{appRFunct}
\mathcal{R}(\Delta\epsilon) & = &
    \sum_{\Tr{nodes}} \sum_{i_1,i_2}
    \int\limits_{-\Lambda}^{0}\dd\epsilon_1 \int\limits_{0}^{\Lambda}\dd\epsilon_2 \hspace{1mm}
    \delta(\epsilon_2 - \epsilon_1 - \Delta\epsilon) | U_{1,2}|^2 \nonumber \\
U_{1,2} & = & \sum_{l_1,l_2} \int\limits_{-\Lambda}^{\Lambda} \dd E_1
    \int\limits_{-\Lambda}^{\Lambda} \dd E_2 \\
& &  \la E_1,l_1 | \epsilon_1,i_1 \ra \la \epsilon_2,i_2 | E_2,l_2 \ra
    \Bf{r}_{\Tr{v}}(\omega) \Bf{U}_{E_1,l_1;E_2,l_2}  \ . \nonumber
\end{eqnarray}
Note that conventionally the summation over fictitious $i_1$ and $i_2$ would be replaced by the appropriate density of states factors. We integrate out $\epsilon_1$ in  ~(\ref{appRFunct}) and write:
\begin{eqnarray}\label{RFunct2}
\mathcal{R}(\Delta\epsilon) & = & \sum_{\Tr{nodes}} \sum_{i_1,i_2}
    \int\limits_{0}^{\Delta\epsilon}\dd\epsilon | U_{1,2} |^2 \nonumber \\
U_{1,2} & = & \sum_{l} \sum_{\sigma=\pm 1}
   \int\limits_{-\Lambda}^{\Lambda} \dd E_1 \int\limits_{-\Lambda}^{\Lambda} \dd E_2 \\
& & \hspace{-10mm}  \Psi^*_{\epsilon-\Delta\epsilon,i_1}(E_1,l-\sigma) \Psi^{\phantom{*}}_{\epsilon,i_2}(E_2,l)
   \times \Bf{r}_{\Tr{v}}(\omega) \Bf{U}_{E_1,l-\sigma;E_2,l} \ . \nonumber
\end{eqnarray}
Now we substitute individual terms from the expression ~(\ref{UExpr}) for transition matrix elements $\Bf{U}_{E_1,l_1;E_2,l_2}$ between the basis states, and check what happens as $\Delta\epsilon \to 0$. We start with potentially the most critical transitions, which involve the zero angular momentum channel.

\subsubsection{Transitions involving the zero angular momentum}

The transition matrix elements that we look first at are given in ~(\ref{UExpr}) for $l = \frac{\sigma+1}{2}$:
\begin{eqnarray}\label{UZero}
& & \hspace{-2mm} \Bf{U}_{E_1,l-\sigma;E_2,l} \propto
  \frac{2 \sigma \hspace{1mm} \Tr{sign}(E_1 E_2)}{\pi} \frac{E_1+E_2}{E_1-E_2} + \\
& &  \frac{C_{\sigma}(q_1,q_2)}{2\pi}
      \left\vert \frac{E_1}{E_2} \right\vert^{\frac{\sigma}{2}}
      \frac{E_1 + E_2}{\sqrt{|E_1 E_2|}}
      \log \left( \frac{|E_1|-|E_2|}{|E_1|+|E_2|} \right)^2 \ . \nonumber
\end{eqnarray}
This expression, appropriate in the thermodynamic limit, appears formally infinite at $E_1=E_2$, and we need to regularize it on a finite sample in order to see the exact effect of this singularity. When the transition matrix elements $\Bf{U}$ are calculated on a finite sample with radius $R$, ~(\ref{UZero}) is replaced by an oscillatory expression $U^{(R)}(E_2-E_1)$, with period $\sim R^{-1}$ and local average exactly given by ~(\ref{UZero}). The discrete energies $E$ (for $|E|R \gg 1$) have spacing equal to the half-period of $U^{(R)}$, so that indeed in the $R \to \infty$ limit the summation of $U^{(R)}$ over energies transforms into the integration over the ``local average'' ~(\ref{UZero}) for every $E_2 \neq E_1$. Exactly at $E_2 = E_1 = E$ it turns out that $U^{(R)} \sim \log(|E|R)$. However, this logarithmic divergence is innocuous, because its contribution to the sum over energies is $R^{-1} \log(|E|R) \to 0$ in the $R \to \infty$ limit. In the following, we will symbolically ignore it in the thermodynamic limit by writing a principal part of ~(\ref{UZero}), and focus on the divergent behavior close to $E_1=E_2$. Note that there are no other divergences in this expression, in particular at $E_1 \to 0$, $E_2 \to 0$ or both.

\begin{figure}[b]
\includegraphics[height=1in]{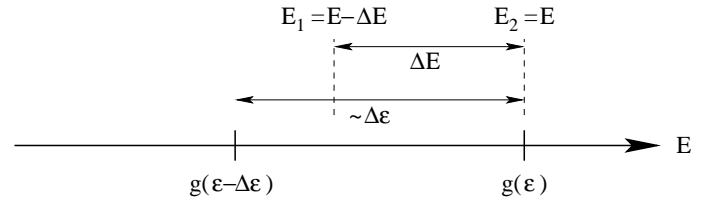}
\caption{\label{Eint}Integration of $U_{1,2}$ in ~(\ref{RFunct2}) over $E$. Ticks mark the points at which the wavefunctions are possibly divergent ($\Psi^*_{\epsilon-\Delta\epsilon,i_1}(E_1,l-\sigma)=\Psi^*_{\epsilon-\Delta\epsilon,i_1}(E-\Delta E,l-\sigma)$ at the left tick, and $\Psi_{\epsilon,i_2}(E_2,l)=\Psi_{\epsilon,i_2}(E,l)$ at the right tick). As we integrate over $E$, for fixed $\Delta E$, the dashed lines move from left to right in this diagram, and cross over the singularities in the wavefunctions. Only for one value of $\Delta E \sim \Delta\epsilon$ both singularities can be crossed at the same time.}
\end{figure}

Consider $U_{1,2}$ in ~(\ref{RFunct2}). We can change variables $E_1 = E - \Delta E$, $E_2 = E$. The transition matrix element ~(\ref{UZero}) diverges only at $\Delta E \to 0$, so let us fix a small finite $\Delta E < \Delta\epsilon$ and imagine integrating out $E$ first in ~(\ref{RFunct2}). For some value of $E$, one of the two wavefunctions $\Psi^*_{\epsilon - \Delta\epsilon,i_1}(E_1,l-\sigma)$ and $\Psi^{\phantom{*}}_{\epsilon,i_2}(E_2,l)$ may become divergent, but not both - see Figure ~\ref{Eint}.
Based on numerics, we assume that a single wavefunction divergence is integrable, so that the integral over $E$ is finite. As a function of $\Delta E$, this integral has a singularity inherited from the expression ~(\ref{UZero}) at $\Delta E = 0$, but its nature has not been changed and no other singularity at any $\Delta E < \Delta\epsilon$ has been introduced. This is important, because if we now proceed with integrating out $\Delta E$, the singularity at $\Delta E = 0$ behaves exactly the same way as if we were simply integrating ~(\ref{UZero}). However, it is not difficult to check that such an integral:
\begin{equation}
\int\limits_{-\Delta E_0}^{\Delta E_0} \dd\Delta E
  \Bigl( \Tr{const.}\frac{\mathbb{P}}{\Delta E} + \Tr{const.} \log ( \Delta E^2 ) \Bigr)
\end{equation}
evaluates to a finite value ($\mathbb{P}$ denotes the principal part due to finite sample regularization).

What remains to be examined is how sending $\Delta\epsilon$ to zero may affect this picture. Generally, the integration over $E$ discussed above picks divergences of both wavefunctions in ~(\ref{RFunct2}), so that the resulting function of $\Delta E$ may be non-analytic or even divergent at some $\Delta E \sim \Delta\epsilon$. In particular, if the only divergences of the wavefunctions are Dirac peaks (as it seems), then the resulting function of $\Delta E$ is divergent as a Dirac peak at $\Delta E = g(\epsilon) - g(\epsilon - \Delta\epsilon) \sim \Delta\epsilon$; other finite non-analyticities in the wavefunctions do not matter. Now, integrating out $\Delta E$ is sensitive to such a divergence and the final result (which should be finite for $\Delta\epsilon \neq 0$) may end up scaling as $\Bf{U}(\Delta E \sim \Delta\epsilon) \sim \frac{1}{\Delta\epsilon}$ for small $\Delta\epsilon$. However, this can actually happen only if we introduce some non-zero chemical potential $\mu$. Otherwise, we always approach the $\Delta\epsilon \sim E_2-E_1 \to 0$ limit only together with $\epsilon \sim E_1+E_2 \to 2\mu = 0$, and no divergence emerges from ~(\ref{RFunct2}) (for details, see analytical calculations in absence of the Doppler shift - this is equivalent to reducing the present exact wavefunctions to pure Dirac peaks, and such divergences do not give rise to dissipation without a non-zero chemical potential).

In conclusion, the transitions that involve the zero angular momentum, which are the most critical for appearance of anomalous vortex dynamics, seem to qualitatively follow the same behavior as if there were no Doppler shift.

\subsubsection{Transitions that do not involve the zero angular momentum}

Consider first the ``most divergent'' terms in ~(\ref{UExpr}) for $l \neq \frac{\sigma+1}{2}$:
\begin{equation}
\Bf{U}_{E_1,l-\sigma;E_2,l} \propto  4 \sqrt{|E_1 E_2|} \delta(E_2 - E_1)
  \Tr{sign} \left( l - \frac{\sigma+1}{2} \right) \ ,
\end{equation}
and substitute them into ~(\ref{RFunct2}). The $\Tr{sign}$ function takes care of the signs for positive or negative $l$ in ~(\ref{UExpr}). Ignoring $\Bf{r}_{\Tr{v}}(\omega)$ and the vector structure of $\Bf{U}$, we find:
\begin{eqnarray}
U_{1,2} & \propto & \sum_{\sigma=\pm 1} \sum_{l}^{l \neq \frac{\sigma+1}{2}} \int\limits_{-\Lambda}^{\Lambda} \dd E
  \Psi^*_{\epsilon-\Delta\epsilon,i_1}(E,l-\sigma) \Psi^{\phantom{*}}_{\epsilon,i_2}(E,l) \nonumber \\
& &  |E| \Tr{sign} \left( l - \frac{\sigma+1}{2} \right) \ .
\end{eqnarray}
According to our assumptions from the beginning of this appendix, the energies $E$ at which the wavefunctions $\Psi_{\epsilon,i}(E,l)$ are non-analytic do not depend on angular momentum $l$. Therefore, we can first sum up $l$ - the result should be finite, and have the same analytical properties as the original expression. Then, we integrate out $E$. As long as $\Delta\epsilon \neq 0$, the two wavefunctions in the integral cannot diverge at the same value of $E$, so that, according to our assumptions, the integral over $E$ is finite. Exactly for $\Delta\epsilon = 0$, the result may be infinite, and it is $\delta( \Delta\epsilon)$ if the wavefunction divergence at $E \sim \epsilon$ is a Dirac peak. This is essentially the same situation as in the case without the Doppler shift: $\delta( \Delta\epsilon)$ is killed by the factors of $\Delta\epsilon$ in the vortex action.

Now consider the remaining portion of ~(\ref{UExpr}) for $l \neq \frac{\sigma+1}{2}$ and substitute:
\begin{eqnarray}
\Bf{U}_{E_1,l-\sigma;E_2,l} & \propto &
   \left( \frac{|E_1|}{|E_2|} \right)^{\sigma l - \frac{1}{2}}
   \frac{E_1 + E_2}{\sqrt{|E_1 E_2|}} \hspace{1mm} \times \nonumber \\
& &   \Theta \left(\Tr{sign} \left( l - \frac{\sigma+1}{2} \right) \sigma (|E_2| - |E_1|) \right)
   \nonumber
\end{eqnarray}
into $U_{1,2}$ from ~(\ref{RFunct2}). The $\Tr{sign}$ function selects whether the argument of the step $\Theta$-function should be $\sigma(k_2-k_1)$ for positive $l$ or $\sigma(k_1-k_2)$ for negative $l$ (see ~(\ref{UExpr}), and recall $k=|E|, q=\Tr{sign}(E)$). We could roughly follow the previous argument, although the summation over $l$ is not entirely trivial - it may introduce a singularity at $E_1 = E_2$. Since the wavefunction amplitudes cannot grow indefinitely with $l$, this new singular behavior is in the worst case:
\begin{equation}
\frac{E_1+E_2}{E_1-E_2}
\end{equation}
(see for example ~(\ref{USumL})). After summation over $l$ we are left with an integral over $E_1$ and $E_2$, which has a very similar structure to the one that we considered in the previous subsection. By repeating the same argument, we arrive to the conclusion that we could obtain anomalous vortex dynamics only with a non-zero chemical potential.

\newpage


\end{document}